\documentclass[proceedings]{JHEP3} % 10pt is ignored!
\PrHEP{PrHEP unesp2002}

\usepackage{epsfig,multicol}                    % please use epsfig.

\newbox\mybox
\newcommand\fverb{\setbox\mybox=\hbox\bgroup\verb}
\newcommand\fverbdo{\egroup\medskip\noindent\fbox{\unhbox\mybox}\ }
\newcommand\fverbit{\egroup\item[\fbox{\unhbox\mybox}]}

\title{Conductance from Non-perturbative Methods I}

\author{Olalla A. Castro-Alvaredo and \speaker{Andreas Fring} \\
       Institut f\"ur Theoretische Physik, Freie Universit\"at Berlin,\\
Arnimallee 14, D-14195 Berlin, Germany \\
        E-mail: \email{Olalla@physik.fu-berlin.de}, \email{Fring@physik.fu-berlin.de }}

\conference{Workshop on Integrable Theories, Solitons and Duality}

\abstract{We investigate different methods to compute the DC conductance in
a quantum wire doped with some impuritied by exploiting  
the integrability of the theories under consideration. 
As an essential ingredient in all methods we 
evaluate the reflection and transmission amplitudes of
the impurities for a variety of defects. When the impurities in
the wire are
coupled to an external three dimensional laser field, we
predict the generation of harmonic emission spectra.
We propose a modified version of the well-known  Kubo formula, 
which incorporates the impurities of the system and evaluate the 
current-current two-point correlation function it involves with 
the help of a form factor expansion. A comparison with the 
corresponding
quantities computed in a Landauer transport theory picture is
carried out in part II. }

\begin{document}

The work I want to report about is based on a series of papers \cite
{CF13,OFC,CF12,CFG,CF10,Fform} with an emphasis on the first two. Olalla
Castro-Alvaredo will present the second part of this talk.

\section{Generalities on conductance}

In the context of 1+1 dimensional quantum field theories an impressive
arsenal of non-perturbative techniques has been developed over the last 25
years. The original motivation was to use the lower dimensional set up as a
testing ground for general conceptual ideas and possibly to apply them in
the context of string theory, such that most of the work in this area can be
characterized very often as rather formal. However, lately the experimental
techniques have advance to such an extent that one might realistically hope
to measure various quantities which can be predicted based on these
approaches.

One of those quantities, which is particularly easy to access, is the
conductance (conductivity). It can be measured in general directly without
perturbing very much the behaviour of the system, e.g. a rigid-lattice bulk
metal, such that the uncertainty of experimental artefacts is reduced to a
minimum. Indeed, there have been some fairly recent measurements \cite{Mill}
of this quantity in 1+1 dimensions and the challenge is of course to explain
these data theoretically and possibly inspire more experiments of a similar
type.

There exist two main theoretical descriptions to compute the conductance,
the Kubo formula \cite{kubo,KTH}, which is the outcome of a dynamical
linear-response theory and the Landauer-B\"{u}ttinger theory \cite{Land},
which is a semi-classical transport theory. The main purpose of the work I
want to present is a comparison between these two descriptions by employing
non-perturbative methods of 1+1 dimensional integrable models. It is in this
sense the wording non-perturbative is to be understood, that is despite the
fact that the overall theoretical description is of a perturbative nature,
within these frameworks we use non-perturbative methods. I will concentrate
on our proposal of a generalized Kubo formula and in the second part,
presented by Olalla Castro-Alvaredo, the computations within the
Landauer-B\"{u}ttinger transport theory framework will be presented.

I will start by anticipating the quantities we have to compute. The system
we consider is a one dimensional quantum wire doped with some impurities
(defects). For the time being we leave the theory describing the wire and
also the nature of the impurities unspecified. In linear response theory one
essentially needs the Fourier transform of the current-current two-point
correlation function. This so-called Kubo formula has been adopted to a
situation with a boundary \cite{LSS}. Since this only captures effects
coming from the constriction of the wire a generalization to a set up with
defects was needed, which we proposed in \cite{CF13} as

\begin{equation}
G^{\mathbf{\alpha }}(T)=-\lim_{\omega \rightarrow 0}\frac{1}{2\omega \pi ^{2}%
}\int\nolimits_{-\infty }^{\infty }dt\,\,e^{i\omega t}\,\,\left\langle
J(t)Z_{\mathbf{\alpha }}\,J(0)\right\rangle _{T,m}.  \label{2}
\end{equation}
Here the defect operator $Z_{\mathbf{\alpha }}$ enters in-between the two
local currents $J$ within the temperature $T$ and mass $m$ dependent
correlation function. The Matsubara frequency is denoted by $\omega $.

The other possibility of determining the conductance which we want to study,
is a generalization of the Landauer-B\"{u}ttinger transport theory picture.
Within this framework a proposal for the conductance through a quantum wire
with a defect (impurity) has been made in \cite{FLS,FLS2} 
\begin{equation}
G^{\mathbf{\alpha }}(T)=\sum_{i}\lim_{(\mu _{i}^{l}-\mu _{i}^{r})\rightarrow
0}\frac{q_{i}}{2}\int\nolimits_{-\infty }^{\infty }d\theta \left[ \rho
_{i}^{r}(\theta ,T,\mu _{i}^{l})|T_{i}^{\mathbf{\alpha }}\left( \theta
\right) |^{2}-\rho _{i}^{r}(\theta ,T,\mu _{i}^{r})|\tilde{T}_{i}^{\mathbf{%
\alpha }}\left( \theta \right) |^{2}\right] ,  \label{1}
\end{equation}
which we only modify to accommodate parity breaking. This means we allow the
transmission amplitudes for a particle of type $i$ with charge $q_{i}$
passing with rapidity $\theta $ through a defect of type $\mathbf{\alpha }$
from the left $T_{i}^{\mathbf{\alpha }}\left( \theta \right) $ and right $%
\tilde{T}_{i}^{\mathbf{\alpha }}\left( \theta \right) $ to be different. The
density distribution function $\rho _{i}^{r}(\theta ,T,\mu _{i})$ depends on
the temperature $T$, and the potential at the left $\mu _{i}^{l}$ and right $%
\mu _{i}^{r}$ constriction of the wire.

The main quantities we have to compute before we can evaluate (\ref{2}) and (%
\ref{1}) are the transmission amplitudes $T_{i}$, the current-current
correlation functions $\langle \ldots \rangle _{T,m}$ and the density
distributions $\rho _{i}$. We obtain all of them non-perturbatively, the $T$%
's by means of potential scattering theory, e.g. \cite{CT}, the correlation
function from a form factor \cite{KW,Smir,BCFK} expansion and the $\rho $'s
from a thermodynamic Bethe (TBA) ansatz \cite{TBAZam} analysis.

\section{Impurity systems}

\subsection{Constraints from the generalized Yang-Baxter equations}

Let me start with the evaluation of the transmission amplitudes, since they
will be required in (\ref{2}) as well as in (\ref{1}). One of the great
advantages of integrability in 1+1 dimensional models is that the n-particle
scattering matrix factorises into two-particle S-matrices, which can be
determined by some constraining equations such as the Yang-Baxter \cite{YB}
and bootstrap equations \cite{boot}. Similar equations hold in the presence
of a boundary \cite{Chered,Skly,FK} or a defect \cite{DMS,CFG}. It is clear
that with regard to the conductance a situation with a pure boundary, i.e.
non-trivial effects on the constrictions, or purely transmitting defects
will be rather uninteresting and we would like to consider the case when $R$
and $T$ are simultaneously non-vanishing. Unfortunately, it will turn out
that for that situation the Yang-Baxter equations are so constraining that
not many integrable theories will be left to consider. Thus this section
serves essentially to motivate the study of the free Fermion, which after
all is very close to a realistic system of electrons propagating in quantum
wires.

We label now particle types by Latin and degrees of freedom of the impurity
by Greek letters, the bulk scattering matrix by $S$, and the left/right
reflection and transmission amplitudes of the defect by $R/\tilde{R}$ and $T/%
\tilde{T}$, respectively. Then the transmission and reflection amplitudes
are constrained by the ``unitarity'' relations 
\begin{eqnarray}
R_{i\alpha }^{j\beta }(\theta )R_{j\beta }^{k\gamma }(-\theta )+T_{i\alpha
}^{j\beta }(\theta )\tilde{T}_{j\beta }^{k\gamma }(-\theta ) &=&\delta
_{i}^{k}\delta _{\alpha }^{\gamma },  \label{U2} \\
R_{i\alpha }^{j\beta }(\theta )T_{j\beta }^{k\gamma }(-\theta )+T_{i\alpha
}^{j\beta }(\theta )\tilde{R}_{j\beta }^{k\gamma }(-\theta ) &=&0\,,
\label{U3}
\end{eqnarray}
and the crossing-hermiticity relations 
\begin{eqnarray}
R_{_{\bar{\jmath}}}^{\alpha }(\theta ) &=&\tilde{R}_{_{\bar{\jmath}%
}}^{\alpha }(-\theta )^{\ast }=S_{j\bar{\jmath}}(2\theta )\tilde{R}%
_{j}^{\alpha }(i\pi -\theta )\,,  \label{c1} \\
T_{_{\bar{\jmath}}}^{\alpha }(\theta ) &=&\tilde{T}_{_{\bar{\jmath}%
}}^{\alpha }(-\theta )^{\ast }=\tilde{T}_{j}^{\alpha }(i\pi -\theta )\,.
\label{c2}
\end{eqnarray}
The equations (\ref{U2}) and (\ref{U3}) also hold after performing a parity
transformation, that is for $R\leftrightarrow \tilde{R}$ and $%
T\leftrightarrow \tilde{T}$.

Depending now on the choice of the initial asymptotic condition one can
derive the following two non-equivalent sets of generalized Yang-Baxter
equations by exploiting the associativity of the extended
Zamolodchikov-Faddeev algebra \cite{Chered,Skly,FK,DMS,CFG} 
\begin{eqnarray}
S(\theta _{12})[\Bbb{I}\otimes R_{\alpha }^{\beta }(\theta _{1})]S(\hat{%
\theta}_{12})[\Bbb{I}\otimes R_{\beta }^{\gamma }(\theta _{2})] &=&[\Bbb{I}%
\otimes R_{\alpha }^{\beta }(\theta _{2})]S(\hat{\theta}_{12})[\Bbb{I}%
\otimes R_{\beta }^{\gamma }(\theta _{1})]S(\theta _{12}),  \label{YBt1} \\
S(\theta _{12})[\Bbb{I}\otimes R_{\alpha }^{\beta }(\theta _{1})]S(\hat{%
\theta}_{12})[\Bbb{I}\otimes T_{\beta }^{\gamma }(\theta _{2})] &=&R_{\beta
}^{\gamma }(\theta _{1})\otimes T_{\alpha }^{\beta }(\theta _{2}),
\label{YBt2} \\
S(\theta _{12})[T_{\alpha }^{\beta }(\theta _{2})\otimes T_{\beta }^{\gamma
}(\theta _{1})] &=&[T_{\alpha }^{\beta }(\theta _{1})\otimes T_{\beta
}^{\gamma }(\theta _{2})]S(\theta _{12}),  \label{YBt3}
\end{eqnarray}
and 
\begin{eqnarray}
R_{\alpha }^{\beta }(\theta _{1})\otimes \tilde{R}_{\beta }^{\gamma }(\theta
_{2}) &=&R_{\beta }^{\gamma }(\theta _{1})\otimes \tilde{R}_{\alpha }^{\beta
}(\theta _{2}),  \label{RR} \\
\lbrack T_{\alpha }^{\beta }(\theta _{2})\otimes \Bbb{I}]S(\hat{\theta}%
_{12})[\tilde{R}_{\beta }^{\gamma }(\theta _{1})\otimes \Bbb{I}]S(\theta
_{12}) &=&T_{\beta }^{\gamma }(\theta _{2})\otimes \tilde{R}_{\alpha
}^{\beta }(\theta _{1}),  \label{TR} \\
\lbrack \Bbb{I}\otimes \tilde{T}_{\alpha }^{\beta }(\theta _{2})]S(\hat{%
\theta}_{12})[\Bbb{I}\otimes R_{\beta }^{\gamma }(\theta _{1})]S(\theta
_{12}) &=&R_{\alpha }^{\beta }(\theta _{1})\otimes \tilde{T}_{\beta
}^{\gamma }(\theta _{2}),  \label{RT} \\
\lbrack T_{\alpha }^{\beta }(\theta _{1})\otimes \Bbb{I}]S(\hat{\theta}%
_{12})[\tilde{T}_{\beta }^{\gamma }(\theta _{2})\otimes \Bbb{I}] &=&[\Bbb{I}%
\otimes \tilde{T}_{\alpha }^{\beta }(\theta _{2})]S(\hat{\theta}_{12})[\Bbb{I%
}\otimes T_{\beta }^{\gamma }(\theta _{1})].  \label{TT}
\end{eqnarray}
We used here the convention $(A\otimes B)_{ij}^{kl}=A_{i}^{k}B_{j}^{l}$ for
the tensor product and abbreviated the rapidity sum $\hat{\theta}%
_{12}=\theta _{1}+\theta _{2}$ and difference $\theta _{12}=\theta
_{1}-\theta _{2}$. Once again the same equations also hold for $%
R\leftrightarrow \tilde{R}$ and $T\leftrightarrow \tilde{T}$.

Apart from some discrepancies in the indices the equations (\ref{YBt1})-(\ref
{YBt3}) correspond to a more simplified, in the sense that there were no
degrees of freedom in the defect and parity invariance is assumed, set of
equations considered previously in \cite{DMS}. For diagonal scattering it
was argued in \cite{DMS} that one can only have reflection and transmission
simultaneously when $S=\pm 1$. In \cite{CFG} a more general set up which
includes all degrees of freedom was studied. A second set of equations (\ref
{RR})-(\ref{TT}), which is not equivalent to (\ref{YBt1})-(\ref{YBt3}) was
found. It was shown that in the absence of degrees of freedom in the defect
no theory which has a non-diagonal bulk scattering matrix admits
simultaneous reflection and transmission. This result even holds for the
completely general case including degrees of freedom in the defect upon a
mild assumption on the commutativity of $R$ and $T$ in these variables. It
was further shown that besides $S=\pm 1$ also the Federbush model \cite
{Feder} and the generalized coupled Federbush models \cite{Fform} allow for $%
R\neq 0$ and $T\neq 0$.

\subsection{Multiple impurity systems}

The most interest situation in impurity systems arises when instead of a
single one considers multiple defects, since that leads to the occurrence of
resonance phenomena and when the number of defects tends to infinity even to
band structures. Assuming that the distance between the defects is small in
comparison to the length of the wire one can easily construct the
transmission and reflection amplitudes of the multiple defect system from
the knowledge of the corresponding quantities in the single defect system.
For instance for two defects one obtains 
\begin{eqnarray}
T_{i}^{\alpha \beta }(\theta ) &=&\frac{T_{i}^{\alpha }(\theta )T_{i}^{\beta
}(\theta )}{1-R_{i}^{\beta }(\theta )\tilde{R}_{i}^{\alpha }(\theta )}%
,\qquad R_{i}^{\alpha \beta }(\theta )=R_{i}^{\alpha }(\theta )+\frac{%
R_{i}^{\beta }(\theta )T_{i}^{\alpha }(\theta )\tilde{T}_{i}^{\alpha
}(\theta )}{1-R_{i}^{\beta }(\theta )\tilde{R}_{i}^{\alpha }(\theta )},
\label{tr} \\
\tilde{T}_{i}^{\alpha \beta }(\theta ) &=&\frac{\tilde{T}_{i}^{\alpha
}(\theta )\tilde{T}_{i}^{\beta }(\theta )}{1-R_{i}^{\beta }(\theta )\tilde{R}%
_{i}^{\alpha }(\theta )},\qquad \tilde{R}_{i}^{\alpha \beta }(\theta )=%
\tilde{R}_{i}^{\beta }(\theta )+\frac{R_{i}^{\alpha }(\theta )T_{i}^{\beta
}(\theta )\tilde{T}_{i}^{\beta }(\theta )}{1-R_{i}^{\beta }(\theta )\tilde{R}%
_{i}^{\alpha }(\theta )}.  \label{tr2}
\end{eqnarray}
These expressions allow for a direct intuitive understanding, for instance
we note that the term $[1-R_{i}^{\beta }(\theta )\tilde{R}_{i}^{\alpha
}(\theta )]^{-1}=\sum_{n=1}^{\infty }(R_{i}^{\beta }(\theta )\tilde{R}%
_{i}^{\alpha }(\theta ))^{n}$ simply results from the infinite number of
reflections which we have in-between the two defects. This is of course well
known from Fabry-Perot type devices of classical and quantum optics. For the
case $T=\tilde{T},R=\tilde{R}$ the expressions (\ref{tr}) and (\ref{tr2})
coincide with the formulae proposed in \cite{Konik2}. When absorbing the
space dependent phase factor into the defect matrices, the explicit example
presented in \cite{DMS} for the free Fermion perturbed with the energy
operator agree almost for $T=\tilde{T},R=\tilde{R}$ \ with the general
formulae (\ref{tr}). They disagree in the sense that the equality of $%
R_{i}^{\alpha \beta }(\theta )$ and $\tilde{R}_{i}^{\alpha \beta }(\theta )$
does not hold for generic $\alpha ,\beta $ as stated in \cite{DMS}.

It is now straightforward to generalize the expressions for an arbitrary
number of defects, say $n$, in a recursive manner 
\begin{eqnarray}
T_{i}^{\vec{\alpha}}(\theta ) &=&\frac{T_{i}^{\alpha _{1}\ldots \alpha
_{k}}(\theta )T_{i}^{\alpha _{k+1}\ldots \alpha _{n}}(\theta )}{1-\tilde{R}%
_{i}^{\alpha _{1}\ldots \alpha _{k}}(\theta )R_{i}^{\alpha _{k+1}\ldots
\alpha _{n}}(\theta )},\qquad \,\,\qquad \qquad \qquad
\,\,\,\,\,\,\,\,\,\,\,1<k<n\,,  \label{ttr} \\
R_{i}^{\vec{\alpha}}(\theta ) &=&R_{i}^{\alpha _{1}\ldots \alpha
_{k}}(\theta )+\frac{R_{i}^{\alpha _{k+1}\ldots \alpha _{n}}(\theta
)T_{i}^{\alpha _{1}\ldots \alpha _{k}}(\theta )\tilde{T}_{i}^{\alpha
_{1}\ldots \alpha _{k}}(\theta )}{1-\tilde{R}_{i}^{\alpha _{1}\ldots \alpha
_{k}}(\theta )R_{i}^{\alpha _{k+1}\ldots \alpha _{n}}(\theta )},\quad
\,\,1<k<n\,.\,\,  \label{ttr2}
\end{eqnarray}
We encoded here the defect degrees of freedom into the vector $\vec{\alpha}%
\mathbf{=}\{\alpha _{1},\cdots ,\alpha _{n}\}$. Similar expressions also
hold for $\tilde{T}_{i}^{\vec{\alpha}}(\theta )=\tilde{T}_{i}^{\alpha
_{1}\ldots \alpha _{n}}(\theta )$ and $\tilde{R}_{i}^{\vec{\alpha}}(\theta )=%
\tilde{R}_{i}^{\alpha _{1}\ldots \alpha _{n}}(\theta )$.

Alternatively, we can define, in analogy to standard quantum mechanical
methods (see e.g. \cite{CT}), a transmission matrix which takes the particle 
$i$ from one side of the defect of type $\alpha $ to the other 
\begin{equation}
\mathcal{M}_{\alpha }^{i}(\theta )=\left( 
\begin{array}{cc}
T_{i}^{\alpha }(\theta )^{-1} & -R_{i}^{\alpha }(\theta )T_{i}^{\alpha
}(\theta )^{-1} \\ 
-R_{i}^{\alpha }(-\theta )T_{i}^{\alpha }(-\theta )^{-1} & T_{i}^{\alpha
}(-\theta )^{-1}
\end{array}
\right) \,.
\end{equation}
Then alternatively to the recursive way (\ref{ttr}) and (\ref{ttr2}), we can
also compute the multi-defect transmission and reflection amplitudes as 
\begin{equation}
T_{i}^{\vec{\alpha}}(\theta )=\left( \prod_{k=1}^{n}\mathcal{M}_{\alpha
_{k}}^{i}(\theta )\right) _{11}^{-1},\,\,\,\quad R_{i}^{\vec{\alpha}}(\theta
)=-\left( \prod_{k=1}^{n}\mathcal{M}_{\alpha _{k}}^{i}(\theta )\right)
_{12}\left( \prod_{k=1}^{n}\mathcal{M}_{\alpha _{k}}^{i}(\theta )\right)
_{11}^{-1}.  \label{ttr3}
\end{equation}
This formulation has the virtue that it is more suitable for numerical
computations, since it just involves matrix multiplications rather than
recurrence operations. In addition it allows for an elegant analytical
computation of the band structures for $n\rightarrow \infty $, which I will
however not comment upon further in this talk.

\subsection{Constraints from potential scattering theory}

\noindent As we argued in section 2.1., in order to obtain a non-trivial
conductance we are lead to consider free theories, possibly with some exotic
statistics. Trying to be as close as possible to some realistic situation,
i.e. electrons, we consider first the free Fermion, which with a line of
defect was first treated in \cite{Cabra}. Thereafter it has also been
considered in \cite{GZ,DMS} and \cite{Konik} from different points of view.
In \cite{Cabra,GZ,DMS} the defect line was taken to be of the form of the
energy operator and in \cite{Konik} also a perturbation in form of a single
Fermion has been considered. In \cite{CF13} we treated a much wider class of
possible defects.

Let us consider the Lagrangian density for a complex free Fermion $\psi $
with $\ell $ defects\footnote{%
We use the conventions: 
\begin{eqnarray*}
x^{\mu } &=&(x^{0},x^{1}),\qquad p^{\mu }=(m\cosh \theta ,m\sinh \theta
),\quad g^{00}=-g^{11}=\varepsilon ^{01}=-\varepsilon ^{10}=1, \\
\gamma ^{0} &=&\left( 
\begin{array}{cc}
0 & 1 \\ 
1 & 0
\end{array}
\right) ,\quad \quad \gamma ^{1}=\left( 
\begin{array}{cc}
0 & 1 \\ 
-1 & 0
\end{array}
\right) ,\quad \gamma ^{5}=\gamma ^{0}\gamma ^{1},\quad \quad \psi _{\alpha
}=\left( 
\begin{array}{c}
\psi _{\alpha }^{(1)} \\ 
\psi _{\alpha }^{(2)}
\end{array}
\right) ,\quad \bar{\psi}_{\alpha }=\psi _{\alpha }^{\dagger }\gamma ^{0}\,.
\end{eqnarray*}
\par
We adopt relativistic units $1=c=\hbar =m\approx e^{2}137$ as mostly used in
the particle physics context rather than atomic units $1=e=\hbar =m\approx
c/137$ more natural in atomic physics.} 
\begin{equation}
\mathcal{L}=\bar{\psi}(i\gamma ^{\mu }\partial _{\mu }-m)\psi
\,+\sum_{n=1}^{\ell }\mathcal{D}^{\alpha _{n}}(\bar{\psi},\psi ,\partial _{t}%
\bar{\psi},\partial _{t}\psi )\delta (x-x_{n})\,.  \label{LF}
\end{equation}
The defect is described here by the functions $\mathcal{D}^{\alpha _{n}}(%
\bar{\psi},\psi ,\partial _{t}\bar{\psi},\partial _{t}\psi ),$ which we
assume to be linear in the Fermi fields $\bar{\psi}$,$\psi $ and their time
derivatives. We can now proceed in analogy to standard quantum mechanical
potential scattering theory (see also \cite{GZ,DMS,Konik}) and construct the
amplitudes by adequate matching conditions on the field. We consider first a
single defect at the origin which suffices, since multiple defect amplitudes
can be constructed from the single defect ones, according to the arguments
of the previous section. We decompose the fields of the bulk theory as $\psi
(x)=\Theta (x)$ $\psi _{+}(x)+\Theta (-x)$ $\psi _{-}(x)$, with $\Theta (x)$
being the Heavyside unit step function, and substitute this ansatz into the
equations of motion. As a matching condition we read off the factors of the
delta function and hence obtain the constraints 
\begin{equation}
i\gamma ^{1}(\psi _{+}(x)-\psi _{-}(x))|_{x=0}=\left. \frac{\partial 
\mathcal{D}}{\partial \bar{\psi}(x)}\right| _{x=0}\,\,-\left. \frac{\partial 
}{\partial t}\left[ \frac{\partial \mathcal{D}}{\partial (\partial _{t}\bar{%
\psi}(x))}\right] \right| _{x=0}.  \label{bbcon}
\end{equation}
We then use for the left ($-$) and right ($+$) parts of $\psi $ the
well-known Fourier decomposition of the free field 
\begin{equation}
\psi _{j}^{f}(x)=\int \frac{d\theta }{\sqrt{4\pi }}\left( a_{j}(\theta
)u_{j}(\theta )e^{-ip_{j}\cdot x}+a_{_{\bar{\jmath}}}^{\dagger }(\theta
)v_{j}(\theta )e^{ip_{j}\cdot x}\right) \,,\qquad  \label{free}
\end{equation}
with the Weyl spinors 
\begin{equation}
u_{j}(\theta )=-i\gamma ^{5}v_{j}(\theta )=\sqrt{\frac{m_{j}}{2}}\left( 
\begin{array}{c}
e^{-\theta /2} \\ 
e^{\theta /2}
\end{array}
\right) \mathrm{\,\,}  \label{WS}
\end{equation}
and substitute them into the constraint (\ref{bbcon}). Treating the
equations obtained in this manner componentwise, stripping off the
integrals, one can bring them thereafter into the form 
\begin{equation}
a_{j,-}(\theta )=R_{_{j}}(\theta )a_{j,-}(-\theta )+T_{_{j}}(\theta
)a_{j,+}(\theta )\,\,,
\end{equation}
which defines the reflection and transmission amplitudes in an obvious
manner. When parity invariance is broken, the corresponding amplitudes from
the right to the left do not have to be identical and we also have 
\begin{equation}
a_{j,+}(-\theta )=\tilde{T}_{_{j}}(\theta )a_{j,-}(-\theta )+\tilde{R}%
_{j}(\theta )a_{j,+}(\theta )\,.
\end{equation}
The creation and annihilation operators $\,a_{i}^{\dagger }(\theta )$ and $%
a_{i}(\theta )$ satisfy the usual fermionic anti-commutation relations $%
\{a_{i}(\theta _{1}),a_{j}(\theta _{2})\}=0$, $\{a_{i}(\theta
_{1}),a_{j}^{\dagger }(\theta _{2})\}=2\pi \delta _{ij}\delta (\theta _{12})$%
. In this way one may construct the $R$'s and $T$'s for any concrete defect
which is of the generic form as described in (\ref{LF}). After the
construction one may convince oneself that the expressions found this way
indeed satisfy the consistency equations like unitarity (\ref{U2}), (\ref{U3}%
) and crossing (\ref{c1}), (\ref{c2}). Unfortunately the equations (\ref{U2}%
)-(\ref{c2}) can not be employed for the construction, since they are not
restrictive enough by themselves to determine the $R$'s and $T$'s. We
consider now some concrete examples:

\subsubsection{Impurities of Luttinger liquid type $\mathcal{D}(\bar{\protect%
\psi},\protect\psi )=\bar{\protect\psi}(g_{1}+g_{2}\protect\gamma ^{0})%
\protect\psi $}

\vspace{-0.2cm} \noindent Luttinger liquids \cite{Lutt} are of great
interest in condensed matter physics, which is one of the motivations for
our concrete choice of the defect $\mathcal{D}(\bar{\psi},\psi )=\bar{\psi}%
(g_{1}+g_{2}\gamma ^{0})\psi $. When taking the conformal limit of the
defect one obtains an impurity which played a role in this context, see e.g. 
\cite{Aff}, after eliminating the bosonic number counting operator. In the
way outlined above, we compute the related transmission and reflection
amplitudes 
\begin{eqnarray}
R_{j}(\theta ,g_{1},g_{2},-y) &=&\tilde{R}_{j}(\theta ,g_{1},g_{2},y)=\frac{%
4i(g_{2}+g_{1}\cosh \theta )e^{2iym\sinh \theta }}{(4+g_{1}^{2}-g_{2}^{2})%
\sinh \theta -4i(g_{1}+g_{2}\cosh \theta )}\,, \\
R_{\bar{\jmath}}(\theta ,g_{1},g_{2},-y) &=&\tilde{R}_{\bar{\jmath}}(\theta
,g_{1},g_{2},y)=\frac{4i(g_{1}-g_{2}\cosh \theta )e^{-2iym\sinh \theta }}{%
(4+g_{1}^{2}-g_{2}^{2})\sinh \theta -4i(g_{1}-g_{2}\cosh \theta )}\,, \\
T_{j}(\theta ,g_{1},g_{2}) &=&\tilde{T}_{j}(\theta ,g_{1},g_{2})=\frac{%
(4+g_{2}^{2}-g_{1}^{2})\sinh \theta }{(4+g_{1}^{2}-g_{2}^{2})\sinh \theta
-4i(g_{1}+g_{2}\cosh \theta )}\,\,, \\
T_{\bar{\jmath}}(\theta ,g_{1},g_{2}) &=&\tilde{T}_{\bar{\jmath}}(\theta
,g_{1},g_{2})=\frac{(4+g_{2}^{2}-g_{1}^{2})\sinh \theta }{%
(4+g_{1}^{2}-g_{2}^{2})\sinh \theta -4i(g_{1}-g_{2}\cosh \theta )}\,.
\end{eqnarray}
In the limit $\lim_{g_{2}\rightarrow 0}\mathcal{D}(\bar{\psi},\psi )=g_{1}%
\bar{\psi}\psi $, we recover the related results for the $T/\tilde{T}$'s and 
$R/\tilde{R}$'s for the energy defect operator. For this type of defect we
present $|T|^{2}$ and $|R|^{2}$ in figure 1 with varying parameters in order
to illustrate some of the characteristics of these functions.

\FIGURE{\epsfig{file=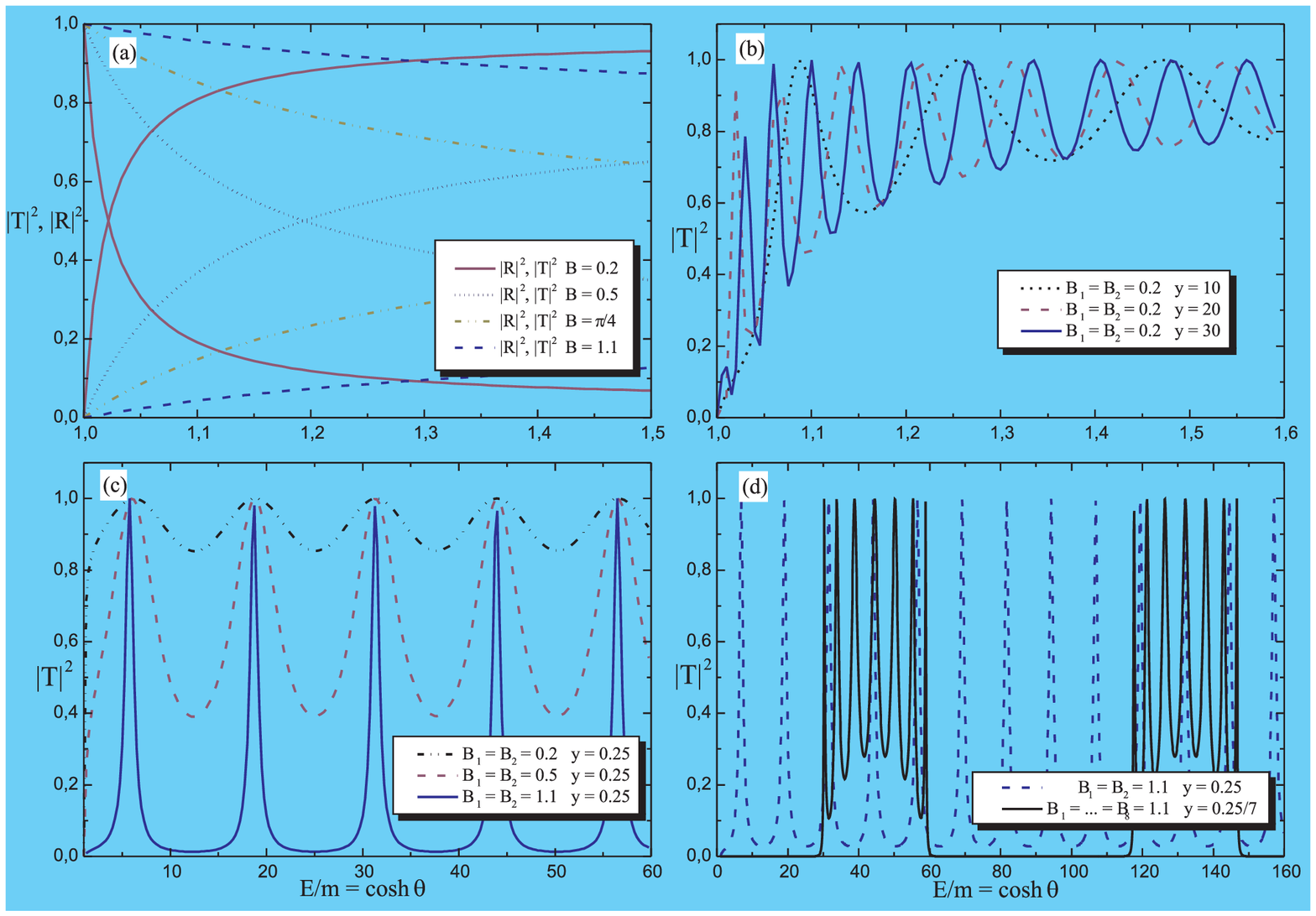,width=15.0cm,height=11.14cm} 
        \caption{  (a) Single defect with varying coupling
constant. $|T|^{2}$ and $|R|^{2}$ correspond to curves starting at 0 and 1
of the same line type, respectively. (b) Double defect with varying distance 
$y$ . (c) Double defect with varying effective coupling constant 
$B=$ arcsin$(-4g_1/(4+g_1^2))  $. 
(d) Double defect $\equiv $  dotted line, eight defects $\equiv $
solid line.}        \label{figure1}}

Part (a) of figure 1 confirms the unitarity relation (\ref{U2}). Part (b)
and (c) show the typical resonances of a double defect, which become
stretched out and pronounced with respect to the energy when the distance
becomes smaller and the coupling constant increases, respectively. Part (d)
exhibits a general feature, that is when the number of defects is increased,
for fixed distance between the outermost defects, the resonances become more
and more dense in that region such that one may speak of energy bands.

\subsubsection{The defect $\mathcal{D}(\bar{\protect\psi},\protect\psi
,\partial _{t}\bar{\protect\psi},\partial _{t}\protect\psi )=ig/2(\bar{%
\protect\psi}\partial _{t}\protect\psi -\partial _{t}\bar{\protect\psi}%
\protect\psi )$}

This type of defect reminds on the first non-trivial charge occurring in the
free Fermion model. In this case we compute by the same means the related
transmission and reflection amplitudes to 
\begin{eqnarray}
\tilde{R}_{j}^{\alpha }(\theta ,y) &=&R_{\bar{\jmath}}^{\alpha }(\theta
,y)=R_{j}^{\alpha }(\theta ,-y)=\tilde{R}_{\bar{\jmath}}^{\alpha }(\theta
,-y)=\frac{-4ig\cosh \theta e^{2iym\sinh \theta }}{4ig+\tanh \theta
(4+g^{2}\cosh ^{2}\theta )}\,,  \label{ren} \\
T_{j}^{\alpha }(\theta ) &=&\tilde{T}_{j}^{\alpha }(\theta )=T_{\bar{\jmath}%
}^{\alpha }(\theta )=\tilde{T}_{\bar{\jmath}}^{\alpha }(\theta )=\frac{%
(4-g^{2}\cosh ^{2}\theta )\tanh \theta }{4ig+\tanh \theta (4+g^{2}\cosh
^{2}\theta )}\,.  \label{ten}
\end{eqnarray}

In \cite{CF13} we also computed the $T/\tilde{T}$'s and $R/\tilde{R}$'s for
other types of defects, such as $\mathcal{D}=g\bar{\psi}\gamma ^{1}\psi $, $%
\mathcal{D}=g\bar{\psi}\gamma ^{5}\psi $, $\mathcal{D}=g\bar{\psi}(\gamma
^{1}\pm \gamma ^{5})\psi \ldots $ As an overall conclusion we observed that
all possible types of parity breaking, that is $T\neq \tilde{T}$; $R\neq 
\tilde{R}$ or $T\neq \tilde{T}$; $R=\tilde{R}$, etc., do occur. We also
confirm a general principle one knows well from quantum mechanics, namely
that parity is preserved when the potential is real, that is in this case
the defect satisfies $\mathcal{D}^{\ast }=\mathcal{D}$.

\subsection{Impurities coupled to laser fields}

Let us now consider a more complex situation in which a three dimensional
laser field hits the quantum wire polarized in such a way that it has a
vector field component along the wire. Since the work of Weyl \cite{Weyl},
one knows that matter may be coupled to light by means of a local gauge
transformation, which reflects itself in the usual minimal coupling
prescription, i.e. $\partial _{\mu }\rightarrow \partial _{\mu }-ieA_{\mu }$%
, with $A_{\mu }$ being the vector gauge potential. The free Fermions in the
wire are then described by the Lagrangian density 
\begin{equation}
\mathcal{L}_{A}=\bar{\psi}(i\gamma ^{\mu }\partial _{\mu }-m+e\gamma ^{\mu
}A_{\mu })\psi \,.  \label{La}
\end{equation}
When the laser field is switched on, we can solve the equation of motion
associated to (\ref{La}) 
\begin{equation}
(i\gamma ^{\mu }\partial _{\mu }-m+e\gamma ^{\mu }A_{\mu })\psi =0
\end{equation}
by a Gordon-Volkov type solution \cite{GV} 
\begin{equation}
\psi _{j}^{A}(x,t)=\exp \left[ ie\int^{x}dsA_{1}(s,t)\right] \psi
_{j}^{f}(x,t)=\exp \left[ ie\int^{t}dsA_{0}(x,s)\right] \psi _{j}^{f}(x,t)\,.
\label{lo2}
\end{equation}
Using now a linearly polarized laser field along the direction of the wire,
the vector potential can typically be taken in the dipole approximation to
be a superposition of monochromatic light with frequency $\omega $, i.e. 
\begin{equation}
A(t):=A_{1}(t)=\frac{1}{x}\int_{0}^{t}dsA_{0}(s)=-\frac{1}{2}%
\int_{0}^{t}dsE(s)=-\frac{E_{0}}{2}\int_{0}^{t}dsf(s)\cos (\omega s)
\label{EF1}
\end{equation}
with $f(t)$ being an arbitrary enveloping function equal to zero for $t<0$
and $t>\tau $, such that $\tau $ denotes the pulse length. In the following
we will always take $f(t)=\Theta (t)\Theta (\tau -t)\,$,$\ $with $\Theta (x)$
being again the Heavyside unit step function. The second equality in (\ref
{EF1}), $A_{0}(x,t)=x\dot{A}(t)$, follows from the fact that we have to
solve (\ref{lo2}).

I want to comment on the validity of the dipole approximation in this
context. It consists usually in neglecting the spatial dependence of the
laser field, which is justified when $x\omega <c=1$, where $x$ is a
representative scale of the problem considered. In the context of atomic
physics this is typically the Bohr radius. In the problem investigated here,
this approximation has to hold over the full spatial range in which the
Fermion follows the electric field. We can estimate this classically, in
which case the maximal amplitude is $eE_{0}/\omega ^{2}$ and therefore the
following constraint has to hold 
\begin{equation}
\left( \frac{eE_{0}}{\omega }\right) ^{2}=4U_{p}<1\,,  \label{Dipole}
\end{equation}
for the dipole approximation to be valid. Due to the fact that $x$ is a
function of $\omega $, we have now a lower bound on the frequency rather
than an upper one as is more common in the context of atomic physics. We
have also introduced here the ponderomotive energy $U_{p}$ for monochromatic
light, that is the average kinetic energy transferred from the laser field
to the electron in the wire.

The solutions to the equations of motion of the free system and the one
which includes the laser field are then related by a factor similar to the
gauge transformation from the length to the velocity gauge 
\begin{equation}
\psi _{j}^{A}(x,t)=\exp \left[ ixeA(t)\right] \psi _{j}^{f}(x)\,.
\label{LLa}
\end{equation}
In an analogous fashion one may use the same minimal coupling procedure also
to couple in addition the laser field to the defect. One has to invoke the
equation of motion in order to carry this out. For convenience we assume now
that the defect is linear in the fields $\bar{\psi}$ and $\psi $. The
Lagrangian density for a complex free Fermion $\psi $ with $\ell $ defects $%
\mathcal{D}^{\alpha }(\bar{\psi},\psi ,A_{\mu })$ of type $\alpha $ at the
position $x_{n}$ subjected to a laser field then reads 
\begin{equation}
\mathcal{L}_{AD}=\mathcal{L}_{A}+\sum_{n=1}^{\ell }\mathcal{D}^{\alpha _{n}}(%
\bar{\psi},\psi ,A_{\mu })\,\delta (x-x_{n})\,.  \label{Lda}
\end{equation}
Considering for simplicity first the case of a single defect situated at $%
x=0 $, the solution to the equation of motion resulting from (\ref{Lda}) is
taken to be of the form $\psi _{j}^{A}(x,t)=\Theta (x)\psi
_{j,+}^{A}(x,t)+\Theta (-x)\psi _{j,-}^{A}(x,t)\,$, which means as before we
distinguish here by notation the solutions (\ref{LLa}) on the left and right
of the defect, $\psi _{j,-}^{A}(x,t)$ and $\psi _{j,+}^{A}(x,t)$,
respectively. Proceeding as before, the matching condition reads now 
\begin{equation}
i\gamma ^{1}(\psi _{j,+}^{A}(x,t)-\psi _{j,-}^{A}(x,t))|_{x=0}=\left. \frac{%
\partial \mathcal{D}_{AD}(\bar{\psi},\psi ,A_{\mu })}{\partial \bar{\psi}%
_{j}^{A}(x,t)}\right| _{x=0}.  \label{bcon}
\end{equation}

\noindent It is clear, that in this case the transmission and reflection
amplitudes will in addition to $\theta $ and $g$ also depend on the
characteristic parameters of the laser field 
\begin{equation}
T(\theta ,g,E_{0},\omega ,t)\quad \quad \mathrm{and\qquad }R(\theta
,g,E_{0},\omega ,t)\,.  \label{TRo}
\end{equation}
With regard to the main theme of this talk, it is clear that the laser field
can be used to control the conductance. For instance defects which have
transmission amplitudes of the form as the solid line in figure 1 (c), can
be used as optically controllable switching devices. I want to deviate now
slightly from the main line of argument and report briefly on an interesting
phenomenon one can predict with solutions of the type (\ref{TRo}).

\subsection{Harmonic generation}

Let me first briefly explain what harmonics are. The first experimental
evidence can be traced back to the early sixties \cite{Franken}. Franken et
al found that when hitting a crystalline quartz with a weak ultraviolet
laser beam of frequency $\omega $, it emits a frequency which is $2\omega $.
Generalizing this phenomenon to higher multiples, one says nowadays that
high harmonics generation is the non-linear response of a medium (a crystal,
an atom, a gas, ...) to a laser field. Harmonic generation is important,
since it allows to convert infrared input radiation of frequency $\omega $
into light in the extreme ultraviolet regime whose frequencies are multiples
of $\omega $ (even up to order $\sim 1000$, see e.g. \cite{record} for a
recent review). A typical experimental spectrum is presented in figure 2.

\smallskip 
\FIGURE{\epsfig{file=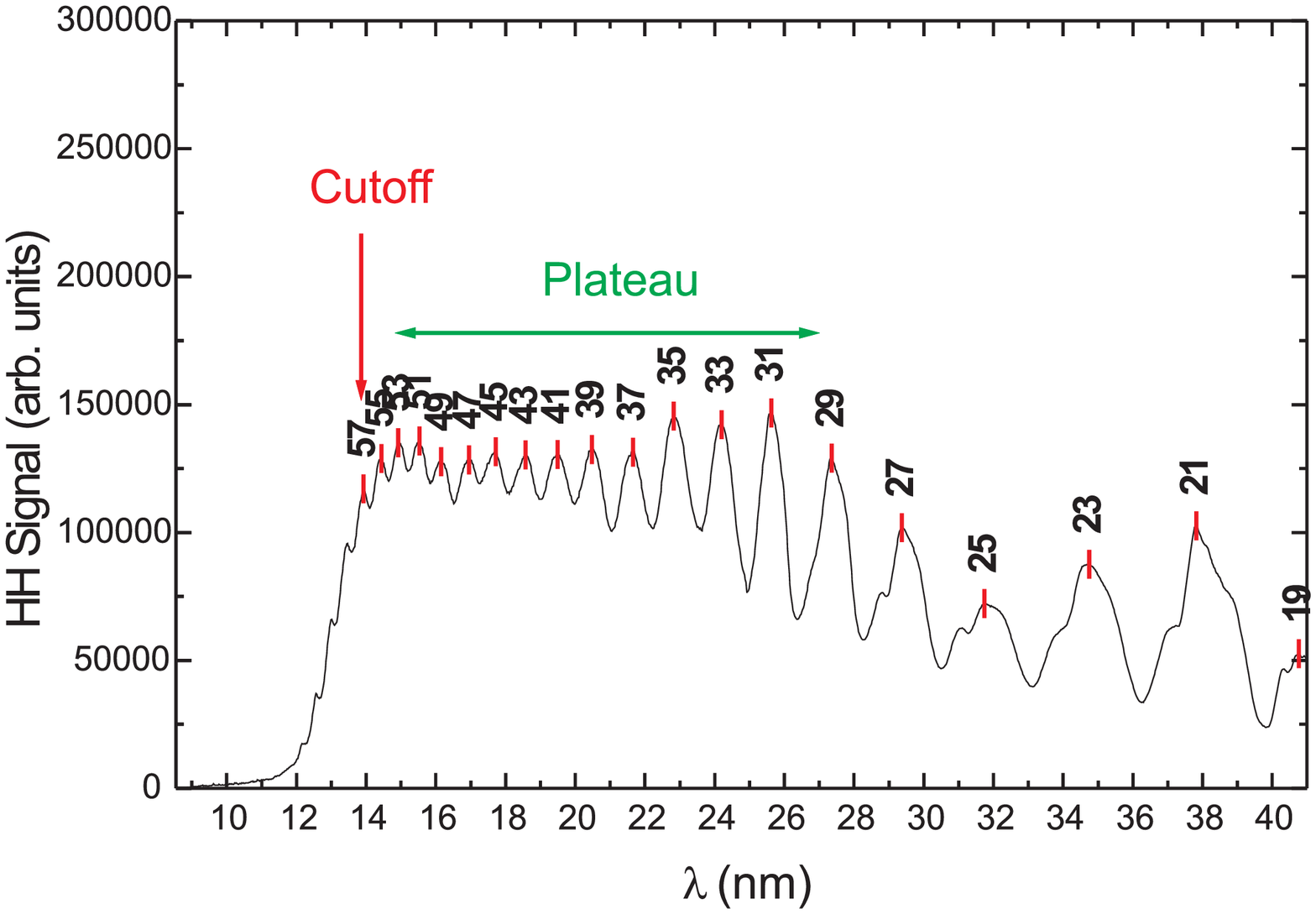,width=8.2cm,height=6.09cm} 
        \caption{Harmonic spectrum for Neon for a Ti:Sa laser with $\lambda= 795nm$. 
Measured at the Max Born Institut Berlin \cite{Phototyp}}    
    \label{figure2}}

In gases, composed of atoms or small molecules, this phenomenon is
well-understood and, to some extent, even controllable in the sense that the
frequency of the highest harmonic, the so-called ``cut-off'', visible in
figure 2, can be tuned as well as the intensities of particular groups of
harmonics. In more complex systems, however, for instance solids, or larger
molecules, high-harmonic generation is still an open problem. This is due to
the fact that, until a few years ago, such systems were expected not to
survive the strong laser fields one needs to produce such effects. However,
nowadays, with the advent of ultrashort pulses, there exist solid-state
materials whose damage threshold is beyond the required intensities of $%
10^{14}\mathrm{W/cm}^{2}$ \cite{solid1}. As a direct consequence, there is
an increasing interest in such materials as potential sources for
high-harmonics. In fact, several groups are currently investigating this
phenomenon in systems such as thin crystals \cite{mois98,thincryst2}, carbon
nanotubes \cite{mois2000}, or organic molecules \cite{benz1,mois2001}.

We will therefore try to answer here the question, whether it is possible to
generate harmonics from solid state devices and as a prototype of such a
system we study a quantum wire coupled to the laser field in the way
described in section 2.4.

In order to answer that question, we first have to study the spectrum of
frequencies which is filtered out by the defect while the laser pulse is
non-zero. The Fourier transforms of the reflection and transmission
probabilities provide exactly this information 
\begin{eqnarray}
\mathcal{T}(\Omega ,\theta ,E_{0},\omega ,\tau ) &=&\frac{1}{\tau }%
\int_{0}^{\tau }dt|T(\theta ,E_{0},\omega ,t)|^{2}\cos (\Omega t),
\label{TTT} \\
\mathcal{R}(\Omega ,\theta ,E_{0},\omega ,\tau ) &=&\frac{1}{\tau }%
\int_{0}^{\tau }dt|R(\theta ,E_{0},\omega ,t)|^{2}\cos (\Omega t).
\end{eqnarray}
When parity is preserved for the reflection and transmission amplitudes,
that is for real defects with $\mathcal{D}^{\ast }=\mathcal{D}$, we have $%
|T|^{2}+|R|^{2}=1$, and it suffices to consider $\mathcal{T}$ \ in the
following.

\subsubsection{Type I defects}

Many features can be understood analytically. Taking the laser field in form
of monochromatic light in the dipole approximation (\ref{EF1}), we may
naturally assume that the transmission probability for some particular
defects can be expanded as 
\begin{equation}
|T_{I}(\theta ,U_{p},\omega ,t)|^{2}=\sum_{k=0}^{\infty }t_{2k}(\theta
)(4U_{p})^{k}\sin ^{2k}(\omega t).  \label{exp}
\end{equation}
We shall refer to defects which admit such an expansion as ``type I
defects''. Assuming that the coefficients $t_{2k}(\theta )$ become at most $%
1 $, we have to restrict our attention to the regime $4U_{p}<1$ in order for
this expansion to be meaningful for all $t$. Note that this is no further
limitation, since it is precisely the same constraint as already encountered
for the validity of the dipole approximation (\ref{Dipole}). The functional
dependence of (\ref{exp}) will turn out to hold for various explicit defects
considered below. Based on this equation, we compute for such type of defect 
\begin{equation}
\mathcal{T}_{I}(\Omega ,\theta ,U_{p},\omega ,\tau )=\sum_{k=0}^{\infty }%
\frac{(2k)!(U_{p})^{k}\sin (\tau \Omega )t_{2k}(\theta )}{\tau \Omega
\prod_{l=1}^{k}[l^{2}-(\Omega /2\omega )^{2}]}\,.  \label{ts}
\end{equation}
It is clear from this expression that type I defects will preferably let
even multiples of the basic frequency $\omega $ pass, whose amplitudes will
depend on the coefficients $t_{2k}(\theta )$. When we choose the pulse
length to be integer cycles, i.e. $\tau =2\pi n/\omega $ for $n\in \Bbb{Z}$,
the expression in (\ref{ts}) reduces even further. The values at even
multiples of the basic frequency are simply 
\begin{equation}
\mathcal{T}_{I}(2n\omega ,\theta ,U_{p})=(-1)^{n}\sum_{k=0}^{\infty
}t_{2k}(\theta )\left( U_{p}\right) ^{k}\left( 
\begin{array}{c}
2k \\ 
k-n
\end{array}
\right) ,  \label{tss}
\end{equation}
which becomes independent of the pulse length $\tau $. Notice also that the
dependence on $E_{0}$ and $\omega $ occurs in the combination of the
ponderomotive energy $U_{p}$. Further statements require the precise form of
the coefficients $t_{2k}(\theta )$ and can only be made with regard to a
more concrete form of the defect.

\subsubsection{Type II defects}

Clearly, not all defects are of the form (\ref{exp}) and we have to consider
also expansions of the type 
\begin{equation}
|T_{II}(\theta ,E_{0}/e,\omega ,t)|^{2}=\sum_{k,p=0}^{\infty
}t_{2k}^{p}(\theta )\frac{E_{0}^{2k+p}}{\omega ^{2k}}\cos ^{p}(\omega t)\sin
^{2k}(\omega t).  \label{exp2}
\end{equation}
We shall refer to defects which admit such an expansion as ``type II
defects''. In this case we obtain 
\begin{eqnarray}
\mathcal{T}_{II}(\Omega ,\theta ,E_{0}/e,\omega ,\tau )
&=&\sum_{k,p=0}^{\infty }\sum_{l=0}^{p}\left( 
\begin{array}{c}
p \\ 
l
\end{array}
\right) \frac{\Omega \sin (\tau \Omega )}{(-1)^{l+1}\tau \omega ^{2+2k}}%
E_{0}^{2k+2p}  \nonumber \\
&&\times \left( \frac{(2k+2l)!t_{2k}^{2p}(\theta )}{\prod%
\limits_{q=0}^{k+l}[(2q)^{2}-(\frac{\Omega }{\omega })^{2}]}+\frac{%
(2k+2l)!t_{2k}^{2p+1}(\theta )E_{0}}{\prod\limits_{q=1}^{k+l+1}[(2q-1)^{2}-(%
\frac{\Omega }{\omega })^{2}]}\right) .
\end{eqnarray}
We observe from this expression that type II defects will filter out all
multiples of $\omega $. For\ the pulse being once again of integer cycle
length, this reduces to 
\begin{equation}
\mathcal{T}_{II}(2n\omega ,\theta ,U_{p},E_{0})=\sum_{k,p=0}^{\infty
}\sum_{l=0}^{p}(-1)^{l+n}\frac{t_{2k}^{2p}(\theta )}{2^{2l-2p}}\left(
U_{p}\right) ^{k+p}E_{0}^{2p}\left( 
\begin{array}{c}
p \\ 
l
\end{array}
\right) \left( 
\begin{array}{c}
2k+2l \\ 
k+l-n
\end{array}
\right)  \label{II1}
\end{equation}
and 
\begin{eqnarray}
\mathcal{T}_{II}((2n-1)\omega ,\theta ,E_{0}/e) &=&\sum_{k,p=0}^{\infty
}\sum_{l=0}^{p}(-1)^{l+n+1}\frac{t_{2k}^{2p+1}(\theta )}{2^{2l-2p+1}}\left(
U_{p}\right) ^{k+p}  \nonumber \\
&&\times \left( 
\begin{array}{c}
p \\ 
l
\end{array}
\right) \frac{(2k+2l)!(2n-1)E_{0}^{2p+1}}{(l+k-n+1)!(l+n+k)!},  \label{II2}
\end{eqnarray}
which are again independent of $\tau $. We observe that in this case we can
not combine the $E_{0}$ and $\omega $ into a $U_{p}$.

\subsubsection{One particle approximation}

In spite of the fact that we are dealing with a quantum field theory, it is
known that a one particle approximation to the Dirac equation is very useful
and physically sensible when the external forces vary only slowly on a scale
of a few Compton wavelengths, see e.g. \cite{IZ}. We may therefore define
the spinor wavefunctions 
\begin{eqnarray}
\Psi _{j,u,\theta }(x,t) &:&=\psi _{j}^{A}(x,t)\frac{\left| a_{j}^{\dagger
}(\theta )\right\rangle }{\sqrt{2\pi ^{2}p_{j}^{0}}}=\frac{e^{-i\vec{p}%
_{j}\cdot \vec{x}}}{\sqrt{2\pi p_{j}^{0}}}u_{j}(\theta ) \\
\Psi _{j,v,\theta }(x,t)^{\dagger } &:&=\psi _{j}^{A}(x,t)^{\dagger }\frac{%
\left| a_{j}^{\dagger }(\theta )\right\rangle }{\sqrt{2\pi ^{2}p_{j}^{0}}}=%
\frac{e^{-i\vec{p}_{j}\cdot \vec{x}}}{\sqrt{2\pi p_{j}^{0}}}v_{j}(\theta
)^{\dagger }\,.
\end{eqnarray}
With the help of these functions we obtain then for the defect system 
\begin{eqnarray}
\mathbf{\Psi }_{i,u,\theta }^{A}(x,t) &:&=\psi _{i}^{A}(x,t)\frac{\left|
a_{i,-}^{\dagger }(\theta )\right\rangle }{\sqrt{2\pi ^{2}p_{i}^{0}}}=\Theta
(-x)\left[ \Psi _{i,u,\theta }(x,t)+\Psi _{i,u,-\theta }(x,t)R_{i}^{\ast
}(\theta )\right]  \nonumber \\
&&+\Theta (x)T_{_{i}}^{\ast }(\theta )\left[ \Psi _{i,u,\theta }(x,t)+\Psi
_{i,u,-\theta }(x,t)\tilde{R}_{_{i}}^{\ast }(-\theta )\right]  \label{fx}
\end{eqnarray}
and the same function with $u\rightarrow v$. Since this expression resembles
a free wave, it can not be normalized properly and we have to localize the
wave in form of a wave packet by multiplying with an additional function, $%
\tilde{g}(p,t)$ in (\ref{free}) and its counterpart $g(x,t)$ in (\ref{fx}),
typically a Gau\ss ian. Then for the function $\mathbf{\Phi }_{i,u,\theta
}^{A}(x,t)=g(x,t)\mathbf{\Psi }_{i,u,\theta }^{A}(x,t)$, we can achieve that 
$\left\| \mathbf{\Phi }\right\| =1$.

\subsubsection{Harmonic spectra}

We are now in the position to determine the emission spectrum for which we
need to compute the absolute value of the Fourier transform of the dipole
moment 
\begin{equation}
\mathcal{X}_{j,u,\theta }(\Omega )=\left| \int_{0}^{\tau }dt\,\left\langle 
\mathbf{\Phi }_{j,u,\theta }^{A}(x,t)^{\dagger }x\mathbf{\Phi }_{j,u,\theta
}^{A}(x,t)\right\rangle \exp i\Omega t\right| \,\,.  \label{33}
\end{equation}
We localize now the wave packet in a region much smaller than the classical
estimate for the maximal amplitude the electron will acquire when following
the laser field. We achieve this with a Gau\ss ian $g(x,t)=\exp
(-x^{2}/\Delta )$, where $\Delta \ll eE_{0}/\omega ^{2}$.

\subsubsection{An example: Impurity of energy operator type}

As mentioned this type of defect, i.e. $\mathcal{D}(\bar{\psi},\psi )=g\bar{%
\psi}\psi (x)$ can be obtained in a limit from the defect discussed in
section 2.3.1. Coupling the vector potential minimally to it yields 
\begin{equation}
\mathcal{D}_{AD}(\bar{\psi},\psi ,A_{\mu })=g\bar{\psi}(1+e/m\gamma ^{\mu
}A_{\mu })\psi \,,
\end{equation}
by invoking the equation of motion. We can now determine the reflection and
transmission amplitudes as outlined above 
\begin{eqnarray}
R_{i}(\theta ,g,A/e,y) &=&\tilde{R}_{i}(\theta ,g,-A/e,-y)=R_{\bar{\imath}%
}(\theta ,g,A/e,-y)=\tilde{R}_{\bar{\imath}}(\theta ,g,-A/e,y)=\quad 
\nonumber \\
&&\frac{[y\dot{A}-\cosh \theta ]e^{-2iy\sinh \theta }}{[1-y\dot{A}\cosh
\theta ]-i\frac{g}{4}[\frac{4}{g^{2}}+1+A^{2}-y^{2}\dot{A}^{2}]\sinh \theta }%
\,.  \label{r1}
\end{eqnarray}
We denoted the differentiation with respect to time by a dot. The
transmission amplitudes turn out to be 
\begin{eqnarray}
T_{i}(\theta ,g,A/e,y) &=&\tilde{T}_{i}(\theta ,g,-A/e,-y)=T_{\bar{\imath}%
}(\theta ,g,-A/e,y)=\tilde{T}_{\bar{\imath}}(\theta ,g,A/e,-y)=\quad 
\nonumber \\
&&\frac{i\left[ 1-y^{2}\dot{A}^{2}+(A-\frac{2i}{g})^{2}\right] \sinh \theta 
}{\frac{4}{g}[1-y\dot{A}\cosh \theta ]-i[\frac{4}{g^{2}}+1+A^{2}-y^{2}\dot{A}%
^{2}]\sinh \theta }\,.  \label{t1}
\end{eqnarray}
Locating the defect at $\ y=0$, the derivative of $A$ does not appear
anymore explicitly in (\ref{r1}) and (\ref{t1}), such that it is clear that
this defect is of type I and admits an expansion of the form (\ref{exp}).
With the explicit expressions (\ref{r1}) and (\ref{t1}) at hand, we can
determine all the coefficients $t_{2k}(\theta )$ in (\ref{exp})
analytically. For this purpose let us first bring the transmission amplitude
into the more symmetric form 
\begin{equation}
\left| T_{i}(\theta ,g,A/e)\right| ^{2}=\frac{\tilde{a}_{0}(\theta
,g)+a_{2}(\theta ,g)A^{2}+a_{4}(\theta ,g)A^{4}}{a_{0}(\theta
,g)+a_{2}(\theta ,g)A^{2}+a_{4}(\theta ,g)A^{4}},  \label{tt}
\end{equation}
with 
\begin{eqnarray}
a_{0}(\theta ,g) &=&16g^{2}+(4+g^{2})^{2}\sinh ^{2}\theta ,\quad \quad 
\tilde{a}_{0}(\theta ,g)=(g^{2}-4)^{2}\sinh ^{2}\theta ,\qquad \\
\,\,a_{2}(\theta ,g) &=&2g^{2}(4+g^{2})\sinh ^{2}\theta ,\quad \quad \qquad
\ a_{4}(\theta ,g)=g^{4}\sinh ^{2}\theta .\qquad
\end{eqnarray}
We can now expand $\left| T(\theta ,g,A)\right| ^{2}$ in powers of the field 
$A(t)$ and identify the coefficients $t_{2k}(\theta ,g)$ in (\ref{exp})
thereafter. To achieve this we simply have to carry out the series expansion
of the denominator in (\ref{tt}). The latter admits the following compact
form 
\begin{equation}
\frac{1}{a_{0}(\theta ,g)+a_{2}(\theta ,g)A^{2}+a_{4}(\theta ,g)A^{4}}%
=\sum_{k=0}^{\infty }c_{2k}(\theta ,g)A^{2k},
\end{equation}
with $c_{0}(\theta ,g)=1/a_{0}(\theta ,g)$ and 
\begin{equation}
c_{2k}(\theta ,g)=-\frac{c_{2k-2}(\theta ,g)a_{2}(\theta ,g)+c_{2k-4}(\theta
,g)a_{4}(\theta ,g)}{a_{0}(\theta ,g)},
\end{equation}
for $k>0$. We understand here that all coefficients $c_{2k}$ with $k<0$ are
vanishing, such that from this formula all the coefficients $c_{2k}$ may be
computed recursively. Hence, by comparing with the series expansion (\ref
{exp}), we find the following closed formula for the coefficients $%
t_{2k}(\theta ,g)$%
\begin{equation}
t_{2k}(\theta ,g)=[\tilde{a}_{0}(\theta ,g)-a_{0}(\theta ,g)]c_{2k}(\theta
,g)\quad k>0.
\end{equation}
The first coefficients then simply read 
\begin{eqnarray}
t_{0}(\theta ,g) &=&\frac{\tilde{a}_{0}(\theta ,g)}{a_{0}(\theta ,g)}%
=|T(\theta ,E_{0}=0)|^{2}, \\
t_{2}(\theta ,g) &=&\frac{a_{2}(\theta ,g)}{a_{0}(\theta ,g)}\left[
1-t_{0}(\theta ,g)\right] =\frac{8g^{4}(4+g^{2})\sinh ^{2}2\theta }{%
(16g^{2}+(4+g^{2})^{2}\sinh ^{2}\theta )^{2}}, \\
t_{4}(\theta ,g) &=&\left[ \frac{a_{4}(\theta ,g)}{a_{2}(\theta ,g)}-\frac{%
a_{2}(\theta ,g)}{a_{0}(\theta ,g)}\right] t_{2}(\theta ,g),
\end{eqnarray}
and so on. It is now clear how to obtain also the higher terms analytically,
but since they are rather cumbersome we do not report them here.

Having computed the coefficients $t_{2k}$, we can evaluate the series (\ref
{ts}) and (\ref{tss}) in principle to any desired order. For some concrete
values of the laser and defect parameters the results of our evaluations are
depicted in figure 3.

\FIGURE{\epsfig{file=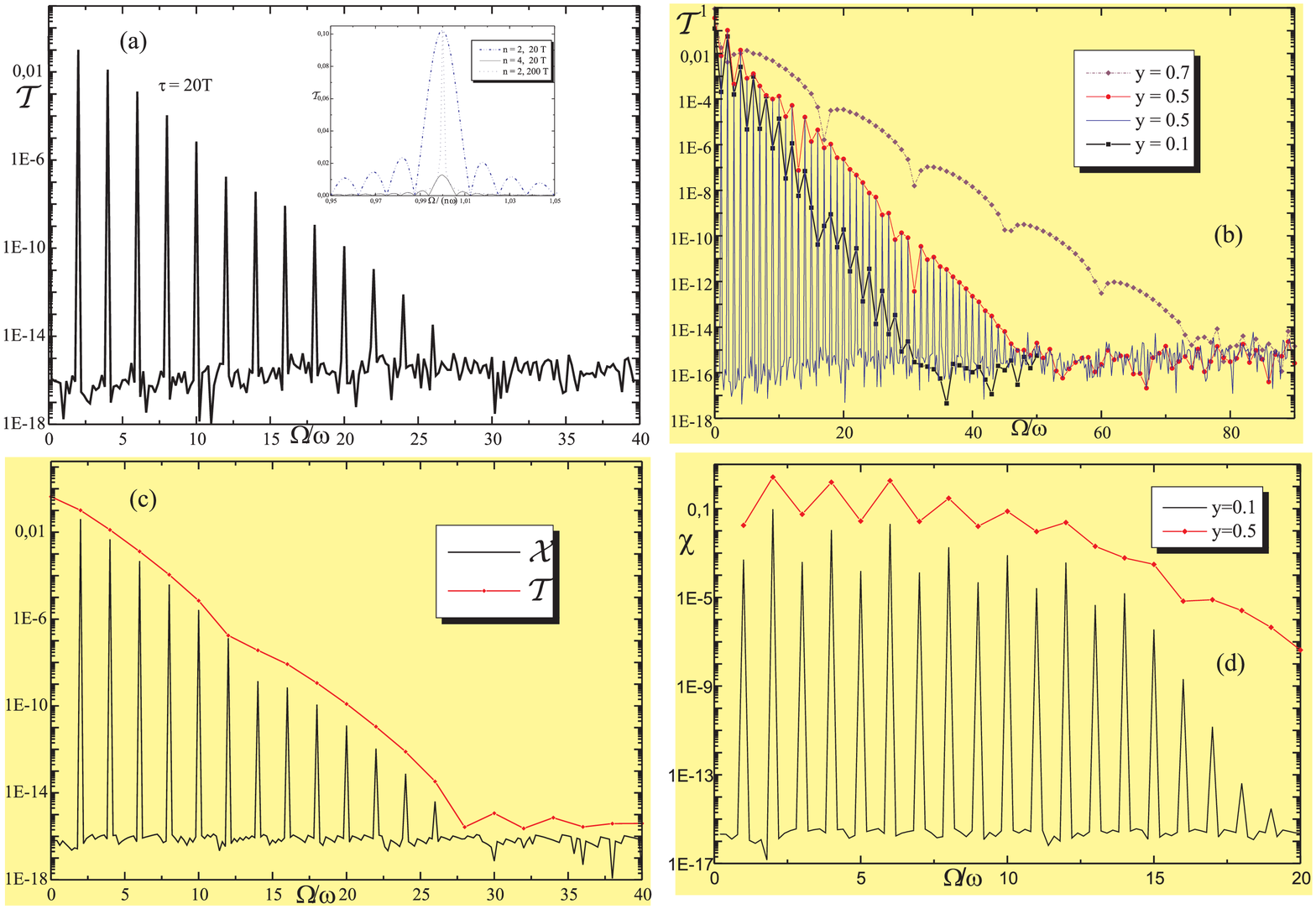,width=15.0cm,height=11.14cm} 
        \caption{Fourier transform of the transmission
probability for a single (a) and double (b) defect with $E_{0}=2.0$, $g=3.5$, 
$\theta =1.2$, $ \omega =0.2$.  Harmonic emission spectrum for a single (c) 
and double (d) defect with $E_{0}=2.0$, $g=3.5$, $\theta =1.2$, $\omega =0.2$, 
$\Delta =6$.  }        \label{figure 3 }}

The main observation from part (a) is that the defect acts as a filter
selecting higher harmonics of even order of the laser frequency.
Furthermore, from the zoom of the peak regions, we see that there are
satellite peaks appearing near the main harmonics. They reduce their
intensity when $\tau $ is increased, such that with longer pulse length the
harmonics become more and more pronounced. We also investigated that for
different frequencies $\omega $ the general structure will not change.
Increasing the field amplitude $E_{0}$, simply lifts up the whole plot
without altering very much its overall structure. We support these findings
in two alternative ways, either by computing directly (\ref{TTT})
numerically or, more instructively, by evaluating the sums (\ref{ts}) and (%
\ref{tss}).

Part (b) shows the analysis for a double defect system with one defect
situated at $x=0$ and the other at $x=y$. The double defect amplitudes are
computed directly from (\ref{tr}) and (\ref{tr2}) with the expression for
the single defect (\ref{r1}) and (\ref{t1}). Since now both $A$ and $\dot{A}$
appear explicitly in the formulae for the $R$'s and $T$'s, it is clear that
the expansion of the double defect can not be of type I, but it turns out to
be of type II, i.e. of the form (\ref{exp2}). Hence, we will now expect that
besides the even also the odd multiples of $\omega $ will be filtered out,
which is indeed visible in part (b) for various distances. Here we have only
plotted a continuous spectrum for $y=0.5$, whereas for reasons of clarity,
we only drew the enveloping function which connects the maxima of the
harmonics for the remaining distances. We observe that now not only odd
multiples of the frequency emerge in addition to the ones in (a) as
harmonics, but also that we obtain much higher harmonics and the cut-off is
shifted further to the ultraviolet. Furthermore, we observe a regular
pattern in the enveloping function, which appears to be independent of $y$.
Similar patterns were observed before in the literature, as for instance in
the context of atomic physics described by a Klein-Gordon formalism (see
figure 2 in \cite{Grob}).

Coming now to the main point of our analysis we would like to see how this
structure is reflected in the harmonic spectra. The result of the evaluation
of (\ref{33}) is depicted in figure 3 parts (c) and (d). We observe a very
similar spectrum as we have already computed for the Fourier transform of
the transmission amplitude, which is not entirely surprising with regard to
the expression (\ref{33}). The cut-off frequencies are essentially
identical. From the comparison between $\mathcal{X}$ and the enveloping
function for $\mathcal{T}$ \ we deduce, that the term involving the
transmission amplitude clearly dominates the spectrum.

The important general deduction from these computations is of course that 
\emph{harmonics of higher order do emerge in the emission spectrum of
impurity systems, such that harmonics can be generated from solid state
devices}.

\section{Conductance from the Kubo formula}

Having characterized various features of defects, I will proceed with the
main theme of the talk, that is the computation of the DC\ conductance. In
the absence of impurities it can be obtained from the Kubo formula in the
form

\begin{equation}
G(T)=-\lim_{\omega \rightarrow 0}\frac{1}{2\omega \pi ^{2}}%
\int\nolimits_{-\infty }^{\infty }dt\,\,e^{i\omega t}\,\,\left\langle
J(t)\,J(0)\right\rangle _{T,m}.  \label{kubo}
\end{equation}
We proposed in \cite{CF13} a generalization of (\ref{kubo}) in the form of (%
\ref{2}). The key quantity needed for the explicit computation of (\ref{kubo}%
) or (\ref{2}) are the occurrence of the temperature dependent
current-current correlation functions $\left\langle J(r)\,J(0)\right\rangle
_{T,m}$ or $\left\langle J(r)\,Z_{\mathbf{\alpha }}J(0)\right\rangle _{T,m}$%
, respectively.

In the zero temperature regime two-point correlation functions can be
computed in general by means of the form factor bootstrap approach \cite
{KW,Smir,BCFK}. In this approach one expands the two-point function between
two local operators $\mathcal{O}$ and $\mathcal{O}^{\prime }$ in terms of
the series

\begin{eqnarray}
\left\langle \mathcal{O}(r)\mathcal{O}^{\prime }(0)\right\rangle _{T=0,m}
&=&\sum\limits_{n=1}^{\infty }\sum\limits_{\mu _{1}\cdots \mu _{n}}\int 
\frac{d\theta _{1}\cdots d\theta _{n}}{n!(2\pi )^{n}}\prod%
\limits_{i=1}^{n}e^{-rm_{i}\cosh \theta _{i}}  \nonumber \\
&&\!\!\!\!\quad \quad \times F_{n}^{\mathcal{O}|\mu _{1}\ldots \mu
_{n}}(\theta _{1},\ldots ,\theta _{n})\left[ F_{n}^{\mathcal{O}^{\prime
}|\mu _{1}\ldots \mu _{n}}(\theta _{1},\ldots ,\theta _{n})\right] ^{\ast },
\label{tzero}
\end{eqnarray}
where we choose $x^{\mu }=(-ir,0)$. The form factors are defined as matrix
elements of the local operator $\mathcal{O}(\vec{x})$ located at the origin
between a multiparticle in-state and the vacuum,

\begin{equation}
F_{n}^{\mathcal{O}|\mu _{1}\ldots \mu _{n}}(\theta _{1},\theta _{2}\ldots
,\theta _{n}):=\langle 0|\mathcal{O}(0)|Z_{\mu _{1}}^{\dagger }(\theta
_{1})Z_{\mu _{2}}^{\dagger }(\theta _{2})\ldots Z_{\mu _{n}}^{\dagger
}(\theta _{n})\rangle .  \label{FF}
\end{equation}
The expansion (\ref{tzero}) is simply obtained by inserting complete states
on the r.h.s. One may proceed similarly by inserting one more set of
complete states when a defect is present and obtains 
\begin{eqnarray}
&&\left\langle J(r)Z_{\mathbf{\alpha }}J(0)\right\rangle
_{T=0,m}\!\!=\!\!\!\sum\limits_{n,m=1}^{\infty }\sum\limits_{\mu _{1}\cdots
\mu _{n};\nu _{1}\cdots \nu _{m}}\int \frac{d\theta _{1}\cdots d\theta _{n}d%
\tilde{\theta}_{1}\cdots d\tilde{\theta}_{m}}{m!n!(2\pi )^{n+m}}F_{n}^{J|\mu
_{1}\ldots \mu _{n}}(\theta _{1}\ldots \theta _{n})  \nonumber \\
&&\!\!\times \left\langle Z_{\mu _{n}}(\theta _{n})\ldots Z_{\mu
_{1}}(\theta _{1})|Z_{\mathbf{\alpha }}|Z_{\nu _{1}}(\tilde{\theta}%
_{1})\ldots Z_{\nu _{m}}(\tilde{\theta}_{m})\right\rangle F_{m}^{J|\nu
_{1}\ldots \nu _{m}}(\tilde{\theta}_{1}\ldots \tilde{\theta}_{m})^{\ast
}e^{-r\sum\limits_{i=1}^{n}m_{i}\cosh \theta _{i}}.\quad \,\,\,\,\,
\label{tree}
\end{eqnarray}
This means there are three principle steps left in order to obtain the
conductance from the expression in (\ref{2}). (a) The computation of the
form factors (\ref{FF}) and the matrix elements involving the defect
operator occurring in (\ref{tree}). (b) The integration in $r$ and (c) the
limit $\omega \rightarrow 0$. Step (a) can be performed in two alternative
ways either by solving certain consistency equations for the form factors
and defect matrix elements or by direct computation. For the latter we
require a representation for the particle creation operators $Z_{\mu
}(\theta )$, the defect operator $Z_{\mathbf{\alpha }}$ and the local
operator $\mathcal{O}(r)$ which is the current in this case.

\subsection{The massless limit}

Remarkably when carrying out the massless limit of the above expressions,
the steps (b) and (c) can be carried out generically. To perform such a
limit we proceed according to the massless limit prescription as suggested
originally in \cite{massless}. It consists of carrying out the limit $%
m\rightarrow 0$ in the high energy regime. In order to do this one replaces
in every rapidity dependent expression $\theta $ by $\theta \pm \sigma $,
where an additional auxiliary parameter $\sigma $ has been introduced.
Thereafter one takes the limit $\sigma \rightarrow \infty $, $m\rightarrow 0$
while keeping the quantity $\hat{m}=m/2\exp (\sigma )$ finite. For instance,
carrying out this prescription for the momentum yields $p_{\pm }=\pm \hat{m}%
\exp (\pm \theta )$, such that one may view the model as splitted into its
two chiral sectors and one can speak naturally of left (L) and right (R)
movers. For the form factors in (\ref{tree}) the massless limit yields 
\begin{equation}
\lim_{\sigma \rightarrow \infty }F_{n}^{\mathcal{O}|\mu _{1}\ldots \mu
_{n}}(\theta _{1}+\eta _{1}\sigma ,\ldots ,\theta _{n}+\eta _{n}\sigma
)=F_{\nu _{1}\cdots \nu _{n}}^{\mathcal{O}|\mu _{1}\ldots \mu _{n}}(\theta
_{1},\ldots ,\theta _{n}),  \label{massff}
\end{equation}
with $\eta _{i}=\pm 1$ and $\nu _{i}=R$ for $\eta _{i}=+$ and $\nu _{i}=L$
for $\eta _{i}=-$. Namely, in the massless limit every massive $n$-particle
form factor is mapped into $2^{n}$ massless form factors. Using these
expressions, performing a Wick rotation and introducing the variable $%
E=\sum_{i=1}^{n}\hat{m}_{i}e^{\theta _{i}}$, we obtain from (\ref{tree}) 
\begin{eqnarray}
&&\left\langle J(r)Z_{\mathbf{\alpha }}J(0)\right\rangle
_{T=m=0}\!\!=\!\!\sum\limits_{n,m=1}^{\infty }\sum\limits_{\mu _{1}\cdots
\mu _{n};\nu _{1}\cdots \nu _{m}}\int \frac{d\theta _{1}\cdots d\theta _{n}d%
\tilde{\theta}_{1}\cdots d\tilde{\theta}_{m}}{m!n!(2\pi )^{n+m}}%
F_{R_{1}\ldots R_{n}}^{J|\mu _{1}\ldots \mu _{n}}(\theta _{1},\ldots ,\theta
_{n})  \nonumber \\
&&\!\!\times \left\langle Z_{\mu _{n}}^{R}(\theta _{n})\ldots Z_{\mu
_{1}}^{R}(\theta _{1})|Z_{\mathbf{\alpha }}|Z_{\nu _{1}}^{R}(\tilde{\theta}%
_{1})\ldots Z_{\nu _{m}}^{R}(\tilde{\theta}_{m})\right\rangle F_{R_{1}\ldots
R_{m}}^{J|\nu _{1}\ldots \nu _{m}}(\tilde{\theta}_{1},\ldots ,\tilde{\theta}%
_{m})^{\ast }e^{-irE}.\quad  \label{tree2}
\end{eqnarray}
We note that for the massless prescription to work, the matrix element
involving the defect $Z_{\mathbf{\alpha }}$ can only depend on the rapidity
differences, which will indeed be the case as we see below. Performing the
variable transformation $\theta _{n}\rightarrow \ln E^{\prime }/\hat{m}%
_{n}-\sum_{i=1}^{n}\hat{m}_{i}/\hat{m}_{n}e^{\theta _{i}}$, we re-write the
r.h.s. of (\ref{tree2}) as 
\begin{eqnarray}
&&\!\!\sum\limits_{n,m=1}^{\infty }\sum\limits_{\mu _{1}\cdots \mu _{n};\nu
_{1}\cdots \nu _{m}}\int_{0}^{E}dE^{\prime }\int\limits_{-\infty }^{\ln
E^{\prime }/\hat{m}_{n}}\frac{d\theta _{1}\cdots d\theta _{n-1}}{n!(2\pi
)^{n}}\int\limits_{-\infty }^{\infty }\frac{d\tilde{\theta}_{1}\cdots d%
\tilde{\theta}_{m}}{m!(2\pi )^{m}}F_{R_{1}\ldots R_{n}}^{J|\mu _{1}\ldots
\mu _{n}}(\theta _{1},\ldots ,\theta _{n}(E^{\prime }))  \nonumber \\
&&\!\!\times \left\langle Z_{\mu _{n}}^{R}(\theta _{n}(E^{\prime }))\ldots
Z_{\mu _{1}}^{R}(\theta _{1})|Z_{\mathbf{\alpha }}|Z_{\nu _{1}}^{R}(\tilde{%
\theta}_{1})\ldots Z_{\nu _{m}}^{R}(\tilde{\theta}_{m})\right\rangle
F_{R_{1}\ldots R_{m}}^{J|\nu _{1}\ldots \nu _{m}}(\tilde{\theta}_{1},\ldots ,%
\tilde{\theta}_{m})^{\ast }e^{-irE^{\prime }}.\quad  \label{Hendrix}
\end{eqnarray}
We substitute now this correlation function into the Kubo formula, shift all
rapidities as $\theta _{i}\rightarrow \theta _{i}+$ $\ln E^{\prime }/\hat{m}%
_{n}$, $\tilde{\theta}_{i}\rightarrow \tilde{\theta}_{i}+$ $\ln E^{\prime }/%
\hat{m}_{n}$, use the Lorentz invariance of the form factors\footnote{%
Denoting by $s$ the Lorentz spin of the operator $\mathcal{O}$ and $\lambda $
being a constant, the form factors satisfy 
\[
F_{n}^{\mathcal{O}|\mu _{1}\ldots \mu _{n}}(\theta _{1}+\lambda ,\ldots
,\theta _{n}+\lambda )=e^{s\lambda }\,F_{n}^{\mathcal{O}|\mu _{1}\ldots \mu
_{n}}(\theta _{1},\ldots ,\theta _{n})\,. 
\]
} and carry out the integration in $dE^{\prime }$%
\begin{eqnarray}
G^{\mathbf{\alpha }}\!\!\! &=&\!\!\!-\lim_{\omega \rightarrow 0}\frac{\omega
^{2s-2}}{\hat{m}_{n}^{2s}\pi }\sum\limits_{\mu _{1}\cdots \mu _{n};\nu
_{1}\cdots \nu _{m}}\int\limits_{-\infty }^{0}\frac{d\theta _{1}\cdots
d\theta _{n-1}}{n!(2\pi )^{n}}\int\limits_{-\infty }^{\infty }\frac{d\tilde{%
\theta}_{1}\cdots d\tilde{\theta}_{m}}{m!(2\pi )^{m}}\frac{1}{%
1-\sum_{i=1}^{n-1}\hat{m}_{i}/\hat{m}_{n}e^{\theta _{i}}}  \nonumber \\
&&\times \left\langle Z_{\mu _{n}}^{R}(\ln (1-\sum\nolimits_{i=1}^{n-1}\hat{m%
}_{i}/\hat{m}_{n}e^{\theta _{i}}))\ldots Z_{\mu _{1}}^{R}(\theta _{1})|Z_{%
\mathbf{\alpha }}|Z_{\nu _{1}}^{R}(\tilde{\theta}_{1})\ldots Z_{\nu
_{m}}^{R}(\tilde{\theta}_{m})\right\rangle \quad  \label{conn} \\
&&\times F_{R_{1}\ldots R_{n}}^{J|\mu _{1}\ldots \mu _{n}}(\theta
_{1},\ldots ,\ln (1-\sum_{i=1}^{n-1}\hat{m}_{i}/\hat{m}_{n}e^{\theta
_{i}}))F_{R_{1}\ldots R_{m}}^{J|\nu _{1}\ldots \nu _{m}}(\tilde{\theta}%
_{1},\ldots ,\tilde{\theta}_{m})^{\ast }\,\,.  \nonumber
\end{eqnarray}
We state various observations: Since the matrix element involving the defect
only depends on the rapidity difference, it is not affected by the shifts.
Operators with Lorentz spin $s=1$ play a very special role in (\ref{conn}),
which makes the current operator especially distinguished. In that case the
r.h.s. of (\ref{conn}) becomes independent of the frequency $\omega $ and
the limit is carried out trivially. Furthermore, since the final expression
has to be independent of $\hat{m}_{n}$, we deduce that the form factors have
to be linearly dependent on $\hat{m}_{n}$.

\subsection{Realization of the defect operator}

\noindent A realization of $Z_{\alpha }$ can be achieved very much in
analogy to a realization of local operators, i.e. as exponentials of
bilinears in Zamolodchikov--Faddeev operators \cite{SMJ}. For the case of a
boundary a generic model independent realization for the boundary operator B
was originally proposed in \cite{GZ} for the parity invariant case, i.e. $R=%
\tilde{R}$ . This proposal was generalized to the defect operator in \cite
{Konik2} with the same restriction and for self-conjugated particles. Here
we extend this realization in order to incorporate the possibility of parity
breaking as well as non self-conjugated particles. A non-trivial consistency
check for the validity of our proposal will be ultimately provided when
exploiting it in the computation of the conductance, obtained by entirely
different means as will be presented in part II. The realization we want to
propose here is a direct generalization of the one presented in \cite{Konik2}%
, namely 
\begin{equation}
Z_{\mathbf{\alpha }}=:\exp [\frac{1}{4\pi }\int\nolimits_{-\infty }^{\infty
}D_{\mathbf{\alpha }}(\theta )\,d\theta ]:\mathrm{{\ },}  \label{D}
\end{equation}
where\ : \ \ : denotes normal ordering and the operator $D_{\alpha }(\theta
) $ has the form 
\begin{eqnarray}
D_{\mathbf{\alpha }}(\theta )\!\!\! &=&\!\!\!\sum\limits_{i}\left[ K_{i}^{%
\mathbf{\alpha }}(\theta )Z_{i}^{\dagger }(\theta )Z_{\bar{\imath}}^{\dagger
}(-\theta )+\tilde{K}_{i}^{\mathbf{\alpha }}(\theta )^{\ast }\,Z_{\bar{\imath%
}}(-\theta )Z_{i}(\theta )\right.  \nonumber \\
&&\left. \qquad +W_{i}^{\mathbf{\alpha }}(\theta )Z_{i}^{\dagger }(\theta
)Z_{i}(\theta )+\tilde{W}_{i}^{\mathbf{\alpha }}(\theta )^{\ast
}Z_{i}^{\dagger }(-\theta )Z_{i}(-\theta )\right] ,  \label{gend}
\end{eqnarray}
with $K_{i}^{\mathbf{\alpha }}(\theta ):=R_{i}^{\mathbf{\alpha }}(\frac{i\pi 
}{2}-\theta )$, $\tilde{K}_{i}^{\mathbf{\alpha }}(\theta ):=\tilde{R}_{i}^{%
\mathbf{\alpha }}(\frac{i\pi }{2}-\theta )$, $W_{i}^{\mathbf{\alpha }%
}(\theta ):=T_{i}^{\mathbf{\alpha }}(\frac{i\pi }{2}-\theta )$ and $\tilde{W}%
_{i}^{\mathbf{\alpha }}(\theta ):=\tilde{T}_{i}^{\mathbf{\alpha }}(\frac{%
i\pi }{2}-\theta )$. In comparison with \cite{Konik2} we have used a
slightly different normalization factor, since in general we have
contributions in the sum over $i$ in (\ref{gend}) including both particles
and anti-particles, as for the complex free Fermion we shall treat below.
Following the arguments given in \cite{GZ}, the operator $D_{\mathbf{\alpha }%
}(\theta )$ depends on the amplitudes $R(\theta )$, $T(\theta )$, $\tilde{R}%
(\theta )$ and $\tilde{T}(\theta )$ with their arguments shifted, as
considered also in \cite{DMS,Konik2}.

\subsection{Defect matrix elements}

\vspace{-0.2cm} \noindent Having now a concrete generic realization of the
defect (\ref{D}), we can compute the defect matrix elements. One way of
doing this is to solve a set of consistency equations which relate the lower
particle matrix elements to higher particle ones, similar as in the standard
form factor program \cite{KW,Smir,BCFK}. Such kind of iterative equations
were proposed in \cite{DMS} for a parity invariant defect and for a real
free fermionic and bosonic theory. We generalize this here and note first
that the operator (\ref{D}) becomes 
\begin{equation}
\lim_{R,\tilde{R}\rightarrow 0;T,\tilde{T}\rightarrow 1}Z_{\mathbf{\alpha }%
}=:\exp [\frac{1}{2\pi }\int\nolimits_{-\infty }^{\infty }d\theta
\sum\limits_{i}Z_{i}^{\dagger }(\theta )Z_{i}(\theta )\,]:,
\end{equation}
and the defect should act in this case as the identity operator, which fixes
our normalization to $\langle Z_{i}(\theta _{1})Z_{\mathbf{\alpha }%
}Z_{j}^{\dagger }(\theta _{2})\rangle =2\pi \,\delta (\theta _{12})\delta
_{ij}$ after having contracted according to Wick's theorem. For two
particles we find, 
\begin{eqnarray}
\left\langle Z_{\bar{\imath}}(\theta _{1})Z_{i}(\theta _{2})Z_{\mathbf{%
\alpha }}\right\rangle &=&\pi \hat{K}_{i}^{\mathbf{\alpha }}(\theta
_{2})\delta (\hat{\theta}_{12}),  \label{i1} \\
\langle Z_{\mathbf{\alpha }}Z_{i}^{\dagger }(\theta _{1})Z_{\bar{\imath}%
}^{\dagger }(\theta _{2})\rangle &=&\pi \,\hat{K}_{i}^{\mathbf{\alpha }%
}(\theta _{1})^{\ast }\delta (\hat{\theta}_{12}),  \label{i2} \\
\langle Z_{i}(\theta _{1})Z_{\mathbf{\alpha }}Z_{j}^{\dagger }(\theta
_{2})\rangle &=&\pi \,\hat{W}_{i}^{\mathbf{\alpha }}(\theta _{1})\delta
(\theta _{12})\delta _{ij}.  \label{i3}
\end{eqnarray}
For later convenience we have introduced the functions 
\begin{eqnarray}
\hat{K}_{i}^{\mathbf{\alpha }}(\theta ) &=&K_{i}^{\mathbf{\alpha }}(\theta
)\,+S_{\bar{\imath}i}(-2\theta )K_{\bar{\imath}}^{\mathbf{\alpha }}(-\theta
)=\tilde{K}_{i}^{\mathbf{\alpha }}(\theta )\,+S_{i\bar{\imath}}(2\theta )%
\tilde{K}_{\bar{\imath}}^{\mathbf{\alpha }}(-\theta ),  \label{uss1} \\
\hat{W}_{i}^{\mathbf{\alpha }}(\theta ) &=&W_{i}^{\mathbf{\alpha }}(\theta
)\,+\tilde{W}_{i}^{\mathbf{\alpha }}(-\theta )^{\ast }=\tilde{W}_{\bar{\imath%
}}^{\mathbf{\alpha }}(-\theta )\,+W_{\bar{\imath}}^{\mathbf{\alpha }}(\theta
)^{\ast }=\hat{W}_{\bar{\imath}}^{\mathbf{\alpha }}(\theta )^{\ast },
\label{uss2}
\end{eqnarray}
since the $K_{i}^{\mathbf{\alpha }}$, $\tilde{K}_{i}^{\mathbf{\alpha }}$,$%
W_{i}^{\mathbf{\alpha }}\,$and $\tilde{W}_{i}^{\mathbf{\alpha }}$ amplitudes
defined before will repeatedly appear in the combinations (\ref{uss1}), (\ref
{uss2}) in what follows. The latter equalities in (\ref{uss1}), (\ref{uss2})
follow simply from 
\begin{equation}
\tilde{W}_{i}^{\mathbf{\alpha }}(\theta )=W_{\bar{\imath}}^{\mathbf{\alpha }%
}(-\theta )=\tilde{W}_{\bar{\imath}}^{\mathbf{\alpha }}(i\pi -\theta )^{\ast
},\mathrm{\,\,}\tilde{K}_{i}^{\mathbf{\alpha }}(\theta )=S_{i\bar{\imath}%
}(2\theta )K_{\bar{\imath}}^{\mathbf{\alpha }}(-\theta )=S_{i\bar{\imath}%
}(2\theta )\tilde{K}_{\bar{\imath}}^{\mathbf{\alpha }}(i\pi -\theta )^{\ast
},  \label{ck}
\end{equation}
which are in turn consequences of the crossing-hermiticity properties (\ref
{c1})-(\ref{c2}). With these matrix elements we can construct the ones
involving more particles recursively from 
\begin{eqnarray}
&&F_{\mathbf{\alpha }}^{\mu _{m}\ldots \mu _{1}\nu _{1}\ldots \nu
_{n}}(\theta _{m}\ldots \theta _{1},\theta _{1}^{\prime }\ldots \theta
_{n}^{\prime }):=\left\langle Z_{\mu _{m}}(\theta _{m})\,\ldots Z_{\mu
_{1}}(\theta _{1})Z_{\mathbf{\alpha }}\,Z_{\nu _{1}}^{\dagger }(\theta
_{1}^{\prime })\ldots Z_{\nu _{n}}^{\dagger }(\theta _{n}^{\prime
})\right\rangle =\quad  \nonumber \\
&&\quad \pi \sum_{l=2}^{m}\delta _{\mu _{1}\bar{\mu}_{l}}\delta (\hat{\theta}%
_{1l})\hat{K}_{\mu _{1}}^{\mathbf{\alpha }}(\theta
_{1})\prod_{p=1}^{l-1}S_{\mu _{1}\mu _{p}}(\theta _{1p})F_{\mathbf{\alpha }%
}^{\mu _{m}\ldots \check{\mu}_{l}\ldots \mu _{2}\nu _{1}\ldots \nu
_{n}}(\theta _{m}\ldots \check{\theta}_{l}\ldots \theta _{2},\theta
_{1}^{\prime }\ldots \theta _{n}^{\prime })  \label{MM} \\
&&+\pi \sum_{l=1}^{n}\delta _{\mu _{1}\nu _{l}}\delta (\theta _{1}-\theta
_{l}^{\prime })\hat{W}_{\mu _{1}}^{\mathbf{\alpha }}(\theta
_{1})\prod_{p=1}^{l-1}S_{\mu _{1}\nu _{p}}(\theta _{1p})F_{\mathbf{\alpha }%
}^{\mu _{m}\ldots \mu _{2}\nu _{1}\ldots \check{\nu}_{l}\ldots \nu
_{n}}(\theta _{m}\ldots \theta _{2},\theta _{1}^{\prime }\ldots \check{\theta%
}_{l}^{\prime }\ldots \theta _{n}^{\prime })  \nonumber
\end{eqnarray}
\begin{eqnarray}
&&F_{\mathbf{\alpha }}^{\mu _{m}\ldots \mu _{1}\nu _{1}\ldots \nu
_{n}}(\theta _{m}\ldots \theta _{1},\theta _{1}^{\prime }\ldots \theta
_{n}^{\prime })=  \label{MM2} \\
&&\quad \pi \sum_{l=2}^{n}\delta _{\nu _{1}\bar{\nu}_{l}}\delta (\hat{\theta}%
_{1l}^{\prime })\hat{K}_{\nu _{1}}^{\mathbf{\alpha }}(\theta _{1}^{\prime
})^{\ast }\prod_{p=1}^{l-1}S_{\nu _{1}\mu _{p}}(\theta _{1p})F_{\mathbf{%
\alpha }}^{\mu _{m}\ldots \mu _{1}\nu _{2}\ldots \check{\nu}_{l}\ldots \nu
_{n}}(\theta _{m}\ldots \theta _{1},\theta _{2}^{\prime }\ldots \check{\theta%
}_{l}^{\prime }\ldots \theta _{n}^{\prime })  \nonumber \\
&&+\pi \sum_{l=1}^{m}\delta _{\nu _{1}\mu _{l}}\delta (\theta _{1}^{\prime
}-\theta _{l})\hat{W}_{\nu _{1}}^{\mathbf{\alpha }}(\theta _{1}^{\prime
})^{\ast }\prod_{p=1}^{l-1}S_{\nu _{1}\mu _{p}}(\theta _{1p})F_{\mathbf{%
\alpha }}^{\mu _{m}\ldots \check{\mu}_{l}\ldots \mu _{1}\nu _{2}\ldots \nu
_{n}}(\theta _{m}\ldots \check{\theta}_{l}\ldots \theta _{1},\theta
_{2}^{\prime }\ldots \theta _{n}^{\prime }).  \nonumber
\end{eqnarray}
Here we denoted with the check on the rapidities $\check{\theta}$ the
absence of the corresponding particle in the matrix element. It is clear
from the expressions (\ref{D}) and (\ref{gend}) that the only possible
non-vanishing matrix elements (\ref{MM}) are those when $n+m$ is even.
Taking (\ref{i1})-(\ref{i3}) as the initial conditions for the recursive
equations (\ref{MM})-(\ref{MM2}), we can now either solve them iteratively
or use (\ref{D}) and evaluate the matrix elements directly. Closed solutions
for these equations have been presented for the first time in \cite{CF13}.

\subsection{Free Fermion wire with impurities}

\vspace{-0.2cm}At this point we have to abandon the general discussion and
consider a concrete theory, which for the reasons already explained we
choose to be the complex free Fermion. Then the generators of the ZF-algebra 
\emph{Z}$_{i}(\theta ),$ \emph{Z}$_{i}^{\dagger }(\theta )$ are just the
usual creation and annihilation operators $a_{i}(\theta ),\,a_{i}^{\dagger
}(\theta )$.

\subsubsection{Defect matrix elements}

\vspace{-0.2cm}Let us now use (\ref{D})-(\ref{gend}) in order to evaluate
matrix elements involving the defect operator. In what follows, the most
relevant matrix elements are those involving four particles, for which we
compute 
\begin{eqnarray*}
\langle a_{i}(\theta _{1})\,a_{\bar{\imath}}(\theta _{2})Z_{\mathbf{\alpha }%
}\,a_{\bar{\imath}}^{\dagger }(\theta _{3})\,a_{i}^{\dagger }(\theta
_{4})\rangle &=&w_{i\bar{\imath}}^{\mathbf{\alpha }}(\theta _{1,}\theta
_{2})\delta (\theta _{14})\delta (\theta _{23})+k_{ii}^{\mathbf{\alpha }%
}(\theta _{1,}\theta _{4})\delta (\hat{\theta}_{12})\delta (\hat{\theta}%
_{34}), \\
\langle a_{i}(\theta _{1})\,a_{i}(\theta _{2})Z_{\mathbf{\alpha }%
}\,a_{j}^{\dagger }(\theta _{3})\,a_{j}^{\dagger }(\theta _{4})\rangle
&=&-\pi ^{2}\hat{W}_{i}^{\mathbf{\alpha }}(\theta _{1})\hat{W}_{i}^{\mathbf{%
\alpha }}(\theta _{2})\delta (\theta _{13})\delta (\theta _{24})\delta _{ij},
\\
\langle a_{i}(\theta _{1})a_{k}(\theta _{2})a_{i}(\theta _{3})Z_{\mathbf{%
\alpha }}a_{i}^{\dagger }(\theta _{4})\rangle &=&\pi ^{2}\hat{W}_{i}^{%
\mathbf{\alpha }}(\theta _{4})\hat{K}_{i}^{\mathbf{\alpha }}(-\theta _{2})%
\left[ \delta (\theta _{14})\delta (\hat{\theta}_{23})-\delta (\hat{\theta}%
_{12})\delta (\theta _{34})\right] \delta _{i\bar{k}}, \\
\langle a_{i}(\theta _{1})Z_{\mathbf{\alpha }}a_{i}^{\dagger }(\theta
_{2})a_{k}^{\dagger }(\theta _{3})a_{i}^{\dagger }(\theta _{4})\rangle
&=&\pi ^{2}\hat{W}_{i}^{\mathbf{\alpha }}(\theta _{1})\hat{K}_{i}^{\mathbf{%
\alpha }}(-\theta _{3})^{\ast }\left[ \delta (\hat{\theta}_{23})\delta
(\theta _{14})-\delta (\theta _{12})\delta (\hat{\theta}_{34})\right] \delta
_{i\bar{k}},
\end{eqnarray*}
with the abbreviations 
\begin{equation}
w_{i\bar{\imath}}^{\mathbf{\alpha }}(\theta _{1,}\theta _{2})=\pi ^{2}\hat{W}%
_{i}^{\mathbf{\alpha }}(\theta _{1})\hat{W}_{\bar{\imath}}^{\mathbf{\alpha }%
}(\theta _{2})\quad \mathrm{and\quad }k_{ii}^{\mathbf{\alpha }}(\theta
_{1,}\theta _{2})=\pi ^{2}\hat{K}_{i}^{\mathbf{\alpha }}(\theta _{1})\hat{K}%
_{i}^{\mathbf{\alpha }}(\theta _{2})^{\ast }\,.  \label{222}
\end{equation}
One can now try to find solutions for all $n$-particle form factors either
from (\ref{MM})-(\ref{MM2}) or by direct computation. For instance for the
stated choice of particles involved, we compute 
\begin{eqnarray}
&&F_{\mathbf{\alpha }}^{m\times (i\bar{\imath})\,n\times (\bar{\imath}%
i)}(\theta _{2m}\ldots \theta _{1},\theta _{1}^{\prime }\ldots \theta
_{2n}^{\prime })=\sum_{k=0}^{\min (n,m)}\frac{(-1)^{m+n-2k}\pi ^{n+m}}{%
(m-k)!(n-k)!k!k!}\int\nolimits_{-\infty }^{\infty }d\beta _{1}\ldots d\beta
_{2n+2m}\,\,  \nonumber \\
&&\times \det \mathcal{A}^{2n}(\beta _{1}\ldots \beta _{2n};\theta
_{1}^{\prime }\ldots \theta _{2n}^{\prime })\det \mathcal{A}^{2m}(\beta
_{2n+1}\ldots \beta _{2n+2m};\theta _{1}\ldots \theta _{2m})  \nonumber \\
&&\times \prod_{p=1}^{k}\hat{W}_{i}^{\mathbf{\alpha }}(\beta _{2p})\hat{W}_{%
\bar{\imath}}^{\mathbf{\alpha }}(\beta _{2p-1})\delta (\beta _{2p}-\beta
_{2n+2p})\delta (\beta _{2p-1}-\beta _{2n+2p-1})  \label{even} \\
&&\times \prod_{p=1+k}^{n}\hat{K}_{i}^{\mathbf{\alpha }}(\beta _{2p})^{\ast
}\delta (\beta _{2p}+\beta _{2p-1})\prod_{p=1+k+n}^{n+m}\hat{K}_{i}^{\mathbf{%
\alpha }}(\beta _{2p})\delta (\beta _{2p}+\beta _{2p-1})\,,  \nonumber
\end{eqnarray}
where $\mathcal{A}^{\ell }(\theta _{1}\ldots \theta _{\ell };\theta
_{1}^{\prime }\ldots \theta _{\ell }^{\prime })$ is a rank $\ell $ matrix
whose entries are given by 
\begin{equation}
\mathcal{A}_{ij}^{\ell }=\cos ^{2}[(i-j)\pi /2]\delta (\theta _{i}-\theta
_{j}^{\prime })\,,\qquad \,\quad 1\leq i,j\leq \ell \,.
\end{equation}
The matrix elements are computed similarly as in \cite{Fform} and references
therein. Likewise we compute 
\begin{eqnarray}
F_{\mathbf{\alpha }}^{n\times i+m\times i}(\theta _{n}\ldots \theta
_{1},\theta _{1}^{\prime }\ldots \theta _{m}^{\prime }) &=&\delta _{n,m}%
\frac{\pi ^{n}(-1)^{n-1}}{n!}\int\nolimits_{-\infty }^{\infty }d\beta
_{1}\ldots d\beta _{n}\prod\limits_{k=1}^{n}\hat{W}_{i}^{\mathbf{\alpha }%
}(\theta _{k})  \nonumber \\
&&\!\!\!\!\!\!\!\!\!\!\!\!\!\!\!\!\!\!\!\!\!\!\!\!\!\!\!\!\!\!\!\!\!\!\!\!\!%
\!\!\!\!\!\!\!\!\!\!\!\!\!\!\!\!\!\!\times \det \mathcal{B}^{n}(\theta
_{n}\ldots \theta _{1};\beta _{1}\ldots \beta _{n})\det \mathcal{B}%
^{n}(\beta _{1}\ldots \beta _{n};\theta _{1}^{\prime }\ldots \theta
_{n}^{\prime }),  \label{allis}
\end{eqnarray}
where we introduced a new rank $\ell $ matrix $\mathcal{B}^{\ell }(\theta
_{1}\ldots \theta _{\ell };\theta _{1}^{\prime }\ldots \theta _{\ell
}^{\prime })$ whose entries are now simply given by 
\begin{equation}
\mathcal{B}_{ij}^{\ell }=\delta (\theta _{i}-\theta _{j}^{\prime }),\qquad
\,\quad 1\leq i,j\leq \ell \,.
\end{equation}
One can verify explicitly \cite{CF13} that these expressions indeed satisfy (%
\ref{MM}) and (\ref{MM2}) .

\subsubsection{Conductance in the $T=m=0$ regime}

\vspace{-0.2cm} \noindent It is well-known that for a free Fermion theory
(also for a single complex free Fermion) the conformal $U(1)$%
-current-current correlation function is simply 
\begin{equation}
\left\langle J(r)J(0)\right\rangle _{T=m=0}=\frac{1}{r^{2}}.  \label{mt0}
\end{equation}
This expression can also be obtained by using the expansion (\ref{tzero}),
together with the massless prescription as outlined above and the
expressions for the only non-vanishing form factors of the current operator
in the complex free Fermion theory 
\begin{equation}
F_{2}^{J|\bar{\imath}i}(\theta ,\tilde{\theta})=-F_{2}^{J|i\bar{\imath}%
}(\theta ,\tilde{\theta})=-i\pi me^{\frac{\theta +\tilde{\theta}}{2}}\,.
\label{currentff}
\end{equation}
In particular, the massless limit of the previous expressions gives,
according to the massless prescription, 
\begin{eqnarray}
F_{RR}^{J|\bar{\imath}i}(\theta ,\tilde{\theta}) &=&-F_{RR}^{J|i\bar{\imath}%
}(\theta ,\tilde{\theta})=-2\pi i\,\hat{m}e^{\frac{\theta +\tilde{\theta}}{2}%
}\,,  \label{mf1} \\
F_{LL}^{J|\bar{\imath}i}(\theta ,\tilde{\theta}) &=&F_{LR}^{J|\bar{\imath}%
i}(\theta ,\tilde{\theta})=F_{RL}^{J|\bar{\imath}i}(\theta ,\tilde{\theta}%
)=F_{LL}^{J|i\bar{\imath}}(\theta ,\tilde{\theta})=F_{LR}^{J|i\bar{\imath}%
}(\theta ,\tilde{\theta})=F_{RL}^{J|i\bar{\imath}}(\theta ,\tilde{\theta}%
)=0\,.\quad  \label{smil}
\end{eqnarray}
We these expressions we can evaluate (\ref{tzero}) to (\ref{mt0}). We may
the insert (\ref{mt0}) into (\ref{kubo}) and the problem is reduced to find
the Fourier transform of the function $r^{-2}$, which is given by $\mathcal{P%
}\int_{-\infty }^{\infty }dr\,\,e^{i\omega r}r^{-2}=-\pi \omega $ for $%
\omega >0$, with $\mathcal{P}$ denoting the principle value. This yields in
the absence of a defect $G(0)=1/2\pi $, in complete agreement with the
well-known classical expression for the conductance in a wire without any
impurities, see for instance \cite{KF}.

For the more complicated situation of $n$ defects $\,Z_{\alpha _{1}}\cdots
Z_{\alpha _{n}}$ located in space at positions $y_{\alpha _{1}}\ldots
y_{\alpha _{n}}$, we compute in the zero temperature and zero mass regime 
\begin{eqnarray}
\left\langle J(r)Z_{\alpha _{1}}\cdots Z_{\alpha _{n}}J(0)\right\rangle
_{T=m=0} &=&\frac{\hat{m}^{2}}{2}\sum\limits_{i}\left[ \int\limits_{-\infty
}^{\infty }\frac{d\theta _{1}}{2}e^{-2r\hat{m}\cosh \theta _{1}}\hat{K}_{i}^{%
\mathbf{\alpha |}R}(\theta _{1})\int\limits_{-\infty }^{\infty }\frac{%
d\theta _{2}}{2}\hat{K}_{i}^{\mathbf{\alpha |}R}(\theta _{2})^{\ast }\right.
\nonumber \\
&&\!\!\!\!\!\!\!\!\!\!\!\!\!\!\!\!\!\!\!\!\!\left. +\int\limits_{-\infty
}^{\infty }\frac{d\theta _{1}}{2}e^{\theta _{1}-r\hat{m}e^{\theta _{1}}}\hat{%
W}_{i}^{\mathbf{\alpha |}R}(\theta _{1})\int\limits_{-\infty }^{\infty }%
\frac{d\theta _{2}}{2}e^{\theta _{2}-r\hat{m}e^{\theta _{2}}}\hat{W}_{\bar{%
\imath}}^{\mathbf{\alpha |}R}(\theta _{2})\right] .  \label{corr}
\end{eqnarray}
The functions $\hat{W}_{i}^{\mathbf{\alpha |}R}(\theta )$, $\hat{K}_{i}^{%
\mathbf{\alpha |}R}(\theta )$, $\ldots $ defined in (\ref{corr}) are the
massless limits of the corresponding functions $\hat{W}_{i}^{\mathbf{\alpha }%
}(\theta )$, $\hat{K}_{i}^{\mathbf{\alpha }}(\theta )$, $\ldots $ For all
the defects we considered, it turned out that the first contribution to the
previous correlation function is actually vanishing, so that (\ref{corr}) is
considerably simplified. In many of the examples, this is due to the fact
that the amplitudes $\hat{K}_{i}^{\mathbf{\alpha }}(\theta )$ are vanishing
in the first place, as a consequence of the crossing relations (\ref{ck}).
The vanishing of the reflection part in (\ref{corr}) also occurs in some
cases as a consequence of the parity of the function $\hat{K}_{i}^{\mathbf{%
\alpha }}(\theta )$. For instance, we find that, for the energy operator
defect such function, although initially non-vanishing, satisfies $\hat{K}%
_{i}^{\mathbf{\alpha }}(\theta )=-\hat{K}_{i}^{\mathbf{\alpha }}(-\theta )$,
such that $\lim_{m\rightarrow 0}\int_{-\infty }^{\infty }d\theta \,\hat{K}%
_{i}^{\mathbf{\alpha }}(\theta )^{\ast }=0$.

\noindent We can now either use (\ref{corr}) to compute the conductance or
evaluate the expression (\ref{conn}) directly in which the frequency limit
is already taken, in both cases we obtain 
\begin{equation}
G^{\mathbf{\alpha }}(0)=\frac{1}{2(2\pi )^{3}}\sum\limits_{i}\int\limits_{-%
\infty }^{0}d\theta \,e^{\theta }\,w_{i\bar{\imath}}^{\mathbf{\alpha |}%
RR}[\ln (1-e^{\theta })_{,}\theta ]\,.
\end{equation}

There are, in addition, further generic results which can be obtained
independently of the specific form of the defect. We present them at this
stage and will confirm their validity below by some specific examples.
Specializing to the case in which all $\ell $ defects are of the same type
and equidistantly separated, i.e. $y=y_{\alpha _{1}}=\cdots =y_{\alpha _{n}}$%
. We can identify two distinct regimes 
\begin{equation}
w_{i\bar{\imath}}^{\mathbf{\alpha |}RR}(\theta _{1,}\theta _{2})=\pi
^{2}\left\{ 
\begin{array}{l}
\overline{\hat{W}_{i}^{\mathbf{\alpha |}R}(\theta _{1})\hat{W}_{i}^{\mathbf{%
\alpha |}R}(\theta _{2})^{\ast }}\quad \,\quad \,\,\,\,\mathrm{%
for\,\,finite\,\,}y \\ 
|\hat{W}_{i}^{\mathbf{\alpha |}R}|^{2}\qquad \qquad \qquad \qquad \,\mathrm{%
for\quad }y\rightarrow 0
\end{array}
\right.  \label{reg}
\end{equation}
where we used in addition (\ref{uss2}). Supported by our explicit examples
below, we find that for $y\rightarrow 0$ in (\ref{reg}) the amplitudes $\hat{%
W}_{i}^{\mathbf{\alpha |}R}(\theta )$ become independent functions of the
rapidity. As we have already argued above 
\begin{equation}
k_{ii}^{\mathbf{\alpha |}RR}(\theta _{1,}\theta _{2})=0.
\end{equation}
It will turn out, that the two regimes specified in (\ref{reg}) are also of
a very distinct nature in the TBA context as presented in part II.

\subsubsection{A wire with impurities of energy operator type}

\noindent Let us exemplify the working of the above formulae with a concrete
defect operator. As a simple example we choose the energy operator defect as
presented in section 2.3.1. Considering first a wire possessing a single
defect of this type, we compute 
\begin{equation}
\hat{W}_{i}^{\alpha }(\theta )=\frac{4\cos B\cosh ^{2}\theta }{\cosh 2\theta
+\cos 2B}\,,\quad \hat{K}_{i}^{\alpha }(\theta )=\frac{2i\sin B\sinh \theta 
}{\sin B-\cosh \theta },\quad w_{i\bar{\imath}}^{\alpha |RR}(\theta
_{1,}\theta _{2})=(2\pi \cos B)^{2}  \label{ggg}
\end{equation}
with $B$ being the effective coupling constant as defined in the caption of
figure 1, such that 
\begin{equation}
\left\langle J(r)Z_{\alpha }J(0)\right\rangle _{T=m=0}=\frac{\cos ^{2}B}{%
r^{2}}\,\,\Longrightarrow \,\,G^{\alpha }(0)=\frac{\cos ^{2}B}{2\pi }.
\label{1d}
\end{equation}
It will turn out that this is in complete agreement with the corresponding
result from the Landauer formula (\ref{2}).

Proceeding in the same way for a wire with two or four impurities we
evaluated \cite{CF13} in the regime $y\gg r$ 
\begin{eqnarray}
\left\langle J(r)Z_{\alpha _{1}}Z_{\alpha _{2}}J(0)\right\rangle _{T=m=0}\!
&=&\frac{4\left[ 1+\sin ^{4}B\right] }{r^{2}\left[ \cos ^{2}(2B)-3\right]
^{2}},  \label{final1} \\
G^{\alpha _{1}\alpha _{2}}(0) &=&\frac{2}{\pi }\!\frac{1+\sin ^{4}B}{\left[
3-\cos ^{2}(2B)\right] ^{2}},  \label{final} \\
\left\langle J(r)Z_{\alpha _{1}}Z_{\alpha _{2}}Z_{\alpha _{3}}Z_{\alpha
_{4}}J(0)\right\rangle _{T=m=0}\! &=&\frac{1}{2r^{2}}\left[ 1+\frac{\cos
^{8}B}{[\cos ^{4}B-2(1+\sin ^{2}B)^{2}]^{2}}\right] , \\
G^{\alpha _{1}\alpha _{2}\alpha _{3}\alpha _{4}}(0) &=&\frac{1}{4\pi }\left(
1+\frac{\cos ^{8}B}{[\cos ^{4}B-2(1+\sin ^{2}B)^{2}]^{2}}\right) \,.
\end{eqnarray}
In the regime $y\rightarrow 0$, we obtained \cite{CF13} 
\begin{eqnarray}
\,\,\lim_{y\rightarrow 0}\left\langle J(r)Z_{\alpha _{1}}Z_{\alpha
_{2}}J(0)\right\rangle _{T=m=0} &=&\frac{1}{r^{2}}\frac{\cos ^{4}B}{(1+\sin
^{2}B)^{2}}\,,\,  \label{y01} \\
\,\,\lim_{y\rightarrow 0}G^{\alpha _{1}\alpha _{2}}(0) &=&\frac{1}{2\pi }%
\frac{\cos ^{4}B}{(1+\sin ^{2}B)^{2}},  \label{y02} \\
\,\lim_{y\rightarrow 0}\left\langle J(r)Z_{\alpha _{1}}Z_{\alpha
_{2}}Z_{\alpha _{3}}Z_{\alpha _{4}}J(0)\right\rangle _{T=m=0} &=&\frac{1}{%
r^{2}}\left( \frac{\cos ^{4}B}{\cos ^{4}B-2(1+\sin ^{2}B)^{2}}\right)
^{2}\,\,,\, \\
\,\lim_{y\rightarrow 0}G^{\alpha _{1}\alpha _{2}\alpha _{3}\alpha _{4}}(0)
&=&\frac{1}{2\pi }\left( \frac{\cos ^{4}B}{\cos ^{4}B-2(1+\sin ^{2}B)^{2}}%
\right) ^{2}\,.
\end{eqnarray}
It will turn out that we can reproduce these expressions by evaluating the
Landauer formula (\ref{1}) when computing the densities with the help of the
TBA. This will now be outlined in part II together with the general
conclusions concerning also this part.

\acknowledgments

We would like to thank the organizers for their kind invitation, financial
support and all their efforts to make this 50$^{\mathrm{th}}$ anniversary
celebration of the Instituto de F\'{i}sica Te\'{o}rica possible. Furthermore
we thank Carla Figueira de Morisson Faria (Max Born Institut Berlin) and
Frank G\"{o}hmann (Universit\"{a}t Bayreuth) for collaboration. We are
grateful to the Deutsche Forschungsgemeinschaft (Sfb288) for financial
support.

\end{document}